\documentclass[journal]{IEEEtran}

\usepackage[T1]{fontenc}
\usepackage[utf8]{inputenc}
\usepackage{graphicx}
\usepackage{amsmath,amssymb,amsfonts}
\usepackage{textcomp}
\usepackage{xcolor}
\usepackage{booktabs}
\usepackage{array}
\usepackage{url}
\usepackage[hidelinks]{hyperref}
\usepackage{cite}
\usepackage{placeins}

\newcolumntype{L}[1]{>{\raggedright\arraybackslash}p{#1}}

\begin{document}

\title{Machine Learning for Specialized QKD Aspects: A Survey of Adaptive Protocols, Free-Space Links, 6G Integration, and Steerability-Aware Security}
\author{Hasan~Abbas~Al-Mohammed and Afnan~S.~Al-Ali%
\thanks{H.~A.~Al-Mohammed and A.~S.~Al-Ali are with Qatar University, Doha, Qatar (e-mail: ha1800217@qu.edu.qa; afnan.alali@qu.edu.qa).}%
}

\maketitle

\begin{abstract}
Quantum Key Distribution (QKD) delivers information-theoretic security
grounded in the laws of quantum mechanics, yet its deployment across
diverse real-world environments goes far beyond conventional
point-to-point fiber links. Several specialized and emerging QKD aspects
are developing rapidly but are rarely treated together: adaptive protocol
and parameter support; free-space, satellite, UAV, and high-altitude
platform (HAP) channels; integration with IoT and 6G networks;
quantum-secured federated learning; QML-assisted decision support; and
steerability-aware estimation for one-sided device-independent QKD. This
survey reviews how Machine Learning (ML), Reinforcement Learning (RL),
and Quantum Machine Learning (QML) address precisely these specialized
scenarios---going beyond conventional point-to-point fiber-link
optimization---and organizes the literature around five thematic pillars:
(I) adaptive protocol and parameter support; (II) free-space, satellite,
UAV, and HAP-assisted QKD; (III) QKD for IoT, 6G, and quantum-secured
federated learning; (IV) QML-assisted QKD functions; and (V)
steerability-aware and one-sided device-independent QKD security
estimation. For each specialized theme we follow a consistent narrative
of problem definition, conventional solution, and ML/RL/QML-based
solution, reporting quantitative gains from the literature using metrics
such as accuracy, mean-absolute-percentage error, QBER reduction, and
SKR improvement. We complement the discussion with per-theme comparison
tables, consolidated cross-theme tables, and a structured analysis of
open challenges specific to these non-terrestrial and application-driven
QKD scenarios, including dataset scarcity for non-standard deployments,
transferability across weather and mobility conditions, interpretability,
trustworthy QML, and the critical boundary between ML as safe
decision-support and ML as security-certification. The survey is intended
as a focused reference for practitioners building adaptive, non-terrestrial,
and application-integrated QKD systems.
\end{abstract}


\section{Introduction}
\IEEEPARstart{T}{he} security of nearly all deployed public-key
cryptography rests on the assumed computational hardness of problems such
as integer factorization and the discrete logarithm~\cite{rivest1978rsa,
diffie1976new}. Shor's algorithm collapses these assumptions on a
sufficiently large fault-tolerant quantum computer~\cite{shor1994algorithms,
grover1996fast}, and the steady experimental progress toward
programmable quantum processors~\cite{arute2019quantum,preskill2018quantum}
has turned the ``harvest-now, decrypt-later'' threat into a concrete
planning concern for long-lived secrets. Two complementary responses have
emerged: post-quantum cryptography, which replaces vulnerable primitives
with conjecturally quantum-resistant ones~\cite{bernstein2017post}, and
Quantum Key Distribution (QKD), which derives the security of the key
itself from physical law rather than computational
assumptions~\cite{bennett1984bb84,ekert1991e91,gisin2002qcrypto}.

QKD enables two parties, conventionally Alice and Bob, to grow a shared
secret key whose secrecy is guaranteed by the no-cloning theorem and the
disturbance that any measurement imparts on non-orthogonal quantum
states~\cite{bennett1984bb84}. Since the seminal BB84
protocol~\cite{bennett1984bb84}, the field has matured into a rich
ecosystem of protocols---entanglement-based~\cite{ekert1991e91},
decoy-state~\cite{hwang2003decoy,lo2005decoy,wang2005decoy},
measurement-device-independent (MDI)~\cite{lo2012mdi}, twin-field
(TF)~\cite{lucamarini2018twinfield}, and continuous-variable
(CV)~\cite{grosshans2002gg02,weedbrook2012gaussian}---supported by
rigorous security proofs~\cite{shor2000simple,mayers2001unconditional,
renner2008security,scarani2009security} and increasingly ambitious field
deployments over fiber and free space~\cite{peev2009secoqc,sasaki2011field,
boaron2018secure,liao2017satellite,yin2020entanglement,chen2021integrated}.
Comprehensive treatments of practical QKD are given in~\cite{scarani2009security,
lo2014secure,xu2020secureqkd,pirandola2020advances}.

\subsection{Motivation: Specialized and Emerging QKD Aspects}
Many existing works study ML for the main QKD operational pipeline---phase
recovery, reconciliation, parameter optimization---and broad cross-layer
surveys~\cite{review_ml_qkd_2024,review_cvqkd_ml_2023} cover this ground
well. However, several important QKD directions are developing separately
and require a focused review that treats them as first-class topics rather
than footnotes. These include: adaptive protocol selection under varying
channel and hardware conditions; free-space and satellite QKD where
atmospheric turbulence, pointing errors, and beam wander dominate; UAV
and high-altitude-platform (HAP) assisted QKD as emerging non-terrestrial
deployment paradigms; QKD integration with IoT and 6G networks under
severe resource and mobility constraints; quantum-secured federated
learning where QKD protects distributed ML; QML-assisted QKD decision
support; and steerability-aware security estimation for one-sided
device-independent QKD. This survey is specifically about these
\emph{specialized and emerging QKD aspects}.

\subsection{What Are Specialized QKD Aspects?}
In this survey, \emph{specialized QKD aspects} refer to QKD functions and
deployment scenarios that go beyond the standard point-to-point fiber-link
pipeline. These include adaptive protocol and parameter support, free-space
and satellite channel characterization, UAV/HAP-assisted QKD, QKD
integration with IoT and 6G systems, quantum-secured federated learning,
QML-assisted QKD decision support, and steerability-aware security
estimation for one-sided device-independent QKD. This definition guides
the entire paper: topics are included when they serve these specialized
themes, and general ML-for-QKD topics (phase recovery, reconciliation,
coexistence, network routing) are discussed only to the extent they support
adaptive, non-terrestrial, or application-driven QKD scenarios.

\subsection{Why Machine Learning for Specialized QKD Aspects?}
The specialized scenarios above share a common difficulty: they are
high-dimensional, nonstationary, and analytically intractable in the forms
that actual deployment demands.
Practical QKD systems must (i) select and adapt protocols and parameters
under finite-key constraints and rapidly changing channel conditions,
especially in mobile and non-terrestrial environments~\cite{lo2005decoy,
wang2019nnparams}; (ii) characterize and compensate fast, nonstationary
atmospheric impairments such as turbulence, beam wander, and pointing
errors in free-space and satellite links~\cite{fried1966optical,
andrews2005laser,vasylyev2012atmospheric}; (iii) operate efficiently over
HAP and UAV platforms where link geometry and weather vary
continuously~\cite{almohammed2025hapxor,almohammed2026hap,
almohammed2024fso_uav}; (iv) deliver QKD-grade security to IoT and 6G
deployments where devices are resource-constrained and distributed~\cite{
almohammed2021icc,almohammed2021iotqkd}; (v) secure federated learning
workflows where key demand is dynamic and unpredictable~\cite{fl_quantum};
and (vi) estimate steerability and support one-sided device-independent
security decisions in near-real-time~\cite{branciard2012onesided,
wiseman2007steering}. These are precisely the regimes where data-driven
Machine Learning (ML) methods excel~\cite{lecun2015deep,jordan2015machine}.

ML has already transformed adjacent fields such as classical optical
communications and networking~\cite{khan2019optical,musumeci2019overview,
oshea2017introduction,zibar2017machine}. Early demonstrations showed that
neural networks can detect attackers during QKD in IoT/B5G networks with
high accuracy~\cite{almohammed2021gcwkshps,almohammed2021access_ml,
almohammed2021iotqkd}, ML can support adaptive QKD post-processing and parameter
control~\cite{almohammed2024cascadeqkd}, and free-space/FSO links can be
secured by combining QKD with adaptive-optics and ML-based channel
estimation~\cite{almohammed2024qkdfso_trains,almohammed2023fso_trains,
almohammed2022fso_tube,almohammed2024fso_uav}. The application of the same
toolbox---random forests~\cite{breiman2001random}, support-vector
machines~\cite{cortes1995support}, gradient-boosted
trees~\cite{chen2016xgboost,ke2017lightgbm}, deep neural
networks~\cite{lecun2015deep,krizhevsky2012imagenet,hochreiter1997lstm},
Bayesian filters~\cite{kalman1960new,julier1997new}, anomaly
detectors~\cite{liu2008isolation,ester1996dbscan}, and (deep) reinforcement
learning~\cite{sutton2018reinforcement,mnih2015human}---to these specialized
QKD aspects is therefore a natural and rapidly growing research frontier.
A further, more speculative direction is native Quantum Machine Learning
(QML), in which the learning model itself runs on quantum
hardware~\cite{biamonte2017quantum,schuld2015introduction,
havlicek2019supervised,dunjko2018machine}.

\subsection{Scope and Contributions}
This survey concentrates on ML, RL and QML applied to \emph{specialized
and emerging QKD aspects}, rather than on the general QKD operational
pipeline or on ML in general. It complements and extends a preliminary
problem-driven survey of classical and ML defenses for
DV/CV-QKD~\cite{almohammed2026survey}. Our contributions are:
\begin{itemize}
\item We define and formalize the concept of \emph{specialized QKD aspects}
in the context of ML-enhanced QKD, establishing a clear scope that
distinguishes this survey from broad ML-for-QKD pipeline surveys.
\item We review ML/RL/QML methods for \emph{adaptive protocol selection
and parameter support}, explaining how learning assists real-time protocol
and parameter decisions beyond conventional static optimization.
\item We survey learning-based methods for \emph{free-space, satellite,
UAV, and HAP-assisted QKD}, covering atmospheric turbulence,
pointing/tracking, mobility, beam wander, weather variation, and
non-terrestrial link scheduling.
\item We review \emph{QKD integration with IoT, 6G, and quantum-secured
federated learning}, emphasizing why these environments create specialized
key distribution requirements and how ML addresses resource constraints,
mobility, and edge intelligence.
\item We discuss \emph{QML-assisted QKD functions}, presenting QML as an
emerging direction with realistic limitations and identifying where quantum
feature maps and variational circuits may offer future benefit.
\item We review \emph{steerability-aware and one-sided device-independent
QKD estimation using ML}, highlighting why this is a distinctive and
practically important specialized security setting.
\item We identify which learning applications are low-risk
decision-support tools and which must be carefully separated from QKD
security proofs, and we summarize open challenges in dataset scarcity,
transferability, interpretability, trustworthy QML, and security-aware
deployment in specialized scenarios.
\end{itemize}

\subsection{Related Surveys and Positioning}
Several recent reviews touch on parts of this landscape. General QKD
networking surveys~\cite{mehic2020quantum,cao2022evolution} treat
architecture and standardization but not learning. The QML literature is
reviewed broadly in~\cite{biamonte2017quantum,dunjko2018machine,
cerezo2022challenges} without a QKD-systems focus. Closer to our scope,
\cite{review_cvqkd_ml_2023} surveys ML specifically for CV-QKD subtasks
(state discrimination, parameter estimation, reconciliation, key-rate
estimation); \cite{review_ml_qkd_2024} organizes ML-for-QKD into parameter
optimization, attack detection, protocol selection, key-performance
prediction and network management; \cite{review_attacks_ml_2024} focuses on
ML for identifying imperfections and attacks; and \cite{review_qml_qkd_2025}
reviews QKD through the lens of QML. Table~\ref{tab:surveys} positions the
present work against these. Our distinguishing features are (a) a focused
thematic scope covering \emph{specialized and emerging QKD aspects} beyond
the conventional fiber-link pipeline, (b) the inclusion of free-space,
satellite, UAV, and HAP-assisted QKD, steerability estimation, and
quantum-secured federated learning as first-class topics, (c) the explicit
problem/classical/ML triad applied uniformly across these specialized themes,
and (d) per-theme and consolidated quantitative tables.

\begin{table*}[!tp]
\centering
\caption{Positioning of this survey relative to recent related reviews across
the coverage dimensions addressed in this work. A tick ($\checkmark$) denotes
full coverage, $\sim$ denotes partial coverage, and --- denotes no coverage.}
\label{tab:surveys}
\begin{tabular}{lccccccc}
\toprule
\textbf{Review} & \textbf{Adapt.} & \textbf{Free-Space} & \textbf{UAV/HAP} &
\textbf{IoT/6G} & \textbf{Fed.\ Learn.} & \textbf{QML} & \textbf{Steer./1SDI} \\
\midrule
Mehic \emph{et al.}~\cite{mehic2020quantum}        & $\sim$ & $\sim$ & --- & $\sim$ & --- & --- & --- \\
Cao \emph{et al.}~\cite{cao2022evolution}          & $\sim$ & $\sim$ & --- & $\sim$ & --- & --- & --- \\
CV-QKD/ML~\cite{review_cvqkd_ml_2023}              & $\checkmark$ & --- & --- & --- & --- & $\sim$ & --- \\
ML-for-QKD~\cite{review_ml_qkd_2024}               & $\checkmark$ & $\sim$ & --- & $\sim$ & $\sim$ & $\sim$ & --- \\
Attacks/ML~\cite{review_attacks_ml_2024}           & --- & --- & --- & $\sim$ & --- & --- & --- \\
QKD via QML~\cite{review_qml_qkd_2025}             & $\sim$ & $\sim$ & --- & $\sim$ & --- & $\checkmark$ & --- \\
Al-Mohammed--Al-Ali~\cite{almohammed2026survey}    & $\sim$ & $\checkmark$ & $\checkmark$ & $\sim$ & --- & $\sim$ & $\sim$ \\
\textbf{This survey}                               & $\checkmark$ & $\checkmark$ & $\checkmark$ & $\checkmark$ & $\checkmark$ & $\checkmark$ & $\checkmark$ \\
\bottomrule
\end{tabular}
\end{table*}
\FloatBarrier

\subsection{Organization}
Section~\ref{sec:background} provides concise QKD and ML background.
Section~\ref{sec:mlmethods} introduces the ML/RL/QML methods used across the
five themes.
Section~\ref{sec:taxonomy} presents the five-theme taxonomy.
Sections~\ref{sec:paramopt}--\ref{sec:steer} cover the five specialized
themes: adaptive protocols; free-space/satellite/UAV/HAP; IoT/6G/federated
learning; QML-assisted functions; and steerability/1SDI-QKD security.
Section~\ref{sec:consolidated} consolidates quantitative results.
Section~\ref{sec:quantitative} gives per-theme gain analysis.
Sections~\ref{sec:datasets}--\ref{sec:verticals} cover datasets, evaluation,
and application deployment.
Section~\ref{sec:synthesis} distills cross-theme design guidelines.
Section~\ref{sec:challenges} presents open challenges and
Section~\ref{sec:roadmap} sets out a ten-item research roadmap.
Section~\ref{sec:conclusion} concludes.

\section{Background}
\label{sec:background}
This section fixes notation and recalls the minimum of QKD and ML needed for
the remainder. Readers familiar with both may skip to
Section~\ref{sec:taxonomy}.

\subsection{Discrete-Variable QKD and the BB84 Protocol}
\begin{figure}[!tb]
\centering
\includegraphics[width=\linewidth]{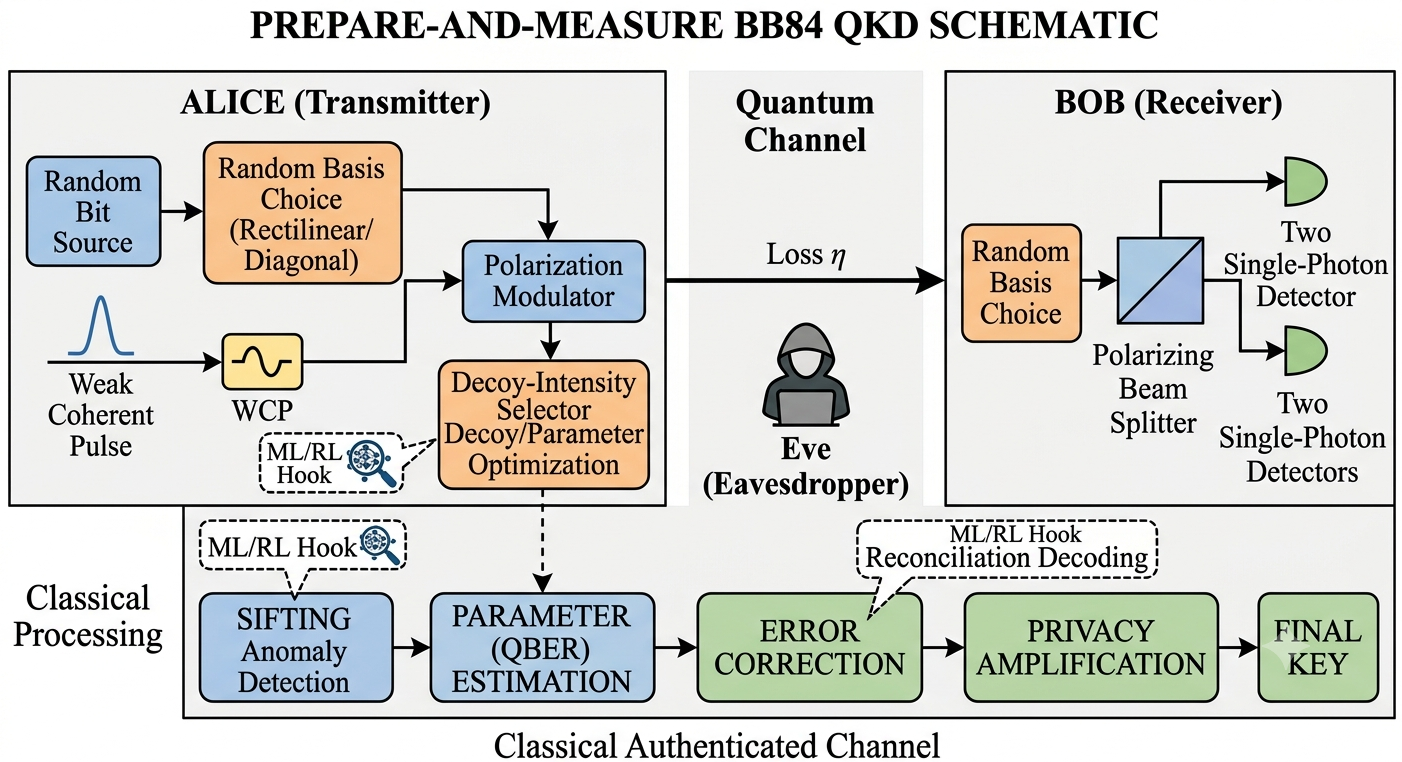}
\caption{Prepare-and-measure BB84 with decoy states, annotated with the
post-processing chain and the ML/RL entry points treated in the survey.}
\label{fig:bb84}
\end{figure}

In BB84~\cite{bennett1984bb84}, Alice encodes random bits in one of two
mutually unbiased bases (e.g.\ rectilinear $\{|0\rangle,|1\rangle\}$ and
diagonal $\{|+\rangle,|-\rangle\}$) and Bob measures in a randomly chosen
basis. After transmission they publicly reconcile bases (\emph{sifting}),
estimate the Quantum Bit Error Rate (QBER) on a sample, perform error
correction (\emph{information reconciliation}) and \emph{privacy
amplification}. For an ideal single-photon BB84 link, the asymptotic secret
fraction is bounded by~\cite{shor2000simple}
\begin{equation}
r \;\geq\; 1 - 2\,h_2(e),
\label{eq:bb84}
\end{equation}
where $e$ is the QBER and $h_2(x)=-x\log_2 x-(1-x)\log_2(1-x)$ is the binary
entropy. Practical weak-coherent-pulse sources are vulnerable to
photon-number-splitting attacks, which the \emph{decoy-state}
method~\cite{hwang2003decoy,lo2005decoy,wang2005decoy} defeats by varying
the mean photon number $\mu$ across signal and decoy intensities and
separately bounding the single-photon yield $Y_1$ and error rate $e_1$. The
decoy-state secret key rate takes the generic form
\begin{equation}
R \;\geq\; q\Big\{ Q_1\big[1-h_2(e_1)\big] - f\,Q_\mu\,h_2(E_\mu) \Big\},
\label{eq:decoy}
\end{equation}
with sifting factor $q$, gains $Q_1,Q_\mu$, overall QBER $E_\mu$, and
reconciliation efficiency $f\!\geq\!1$. The free parameters
$(\mu,\nu,\dots)$ and the probabilities must be \emph{optimized}, often
under finite-key corrections~\cite{scarani2008quantum,tomamichel2012tight};
this optimization is one of the first places ML enters
(Section~\ref{sec:paramopt}). The complete prepare-and-measure chain
and its ML/RL entry points are illustrated in Fig.~\ref{fig:bb84}.

\FloatBarrier

\subsection{Measurement-Device-Independent and Twin-Field QKD}
MDI-QKD~\cite{lo2012mdi} removes all detector side channels by having Alice
and Bob send states to an untrusted relay that performs a Bell-state
measurement; it has been demonstrated over hundreds of
kilometers~\cite{yin2016measurement,comandar2016quantum}. TF-QKD~\cite{lucamarini2018twinfield} encodes information in the phase of weak
coherent fields interfering at a central node, achieving a key-rate scaling
as $O(\sqrt{\eta})$ with channel transmittance $\eta$, thereby surpassing
the repeaterless secret-key capacity (PLOB bound)~\cite{pirandola2017fundamental}; record demonstrations now exceed 500--600~km
\cite{chen2020sending,liu2019experimental,fang2020implementation,
pittaluga2021600}. Both families involve delicate phase and intensity
stabilization that motivate learned controllers. A 2026 experimental
demonstration of COW-QKD achieved information-theoretically secure
transmission to 100~km with kbit/s rates in the finite-key
regime~\cite{experimental_cow_2026}.

\subsection{Continuous-Variable QKD}
\begin{figure}[!tb]
\centering
\includegraphics[width=\linewidth]{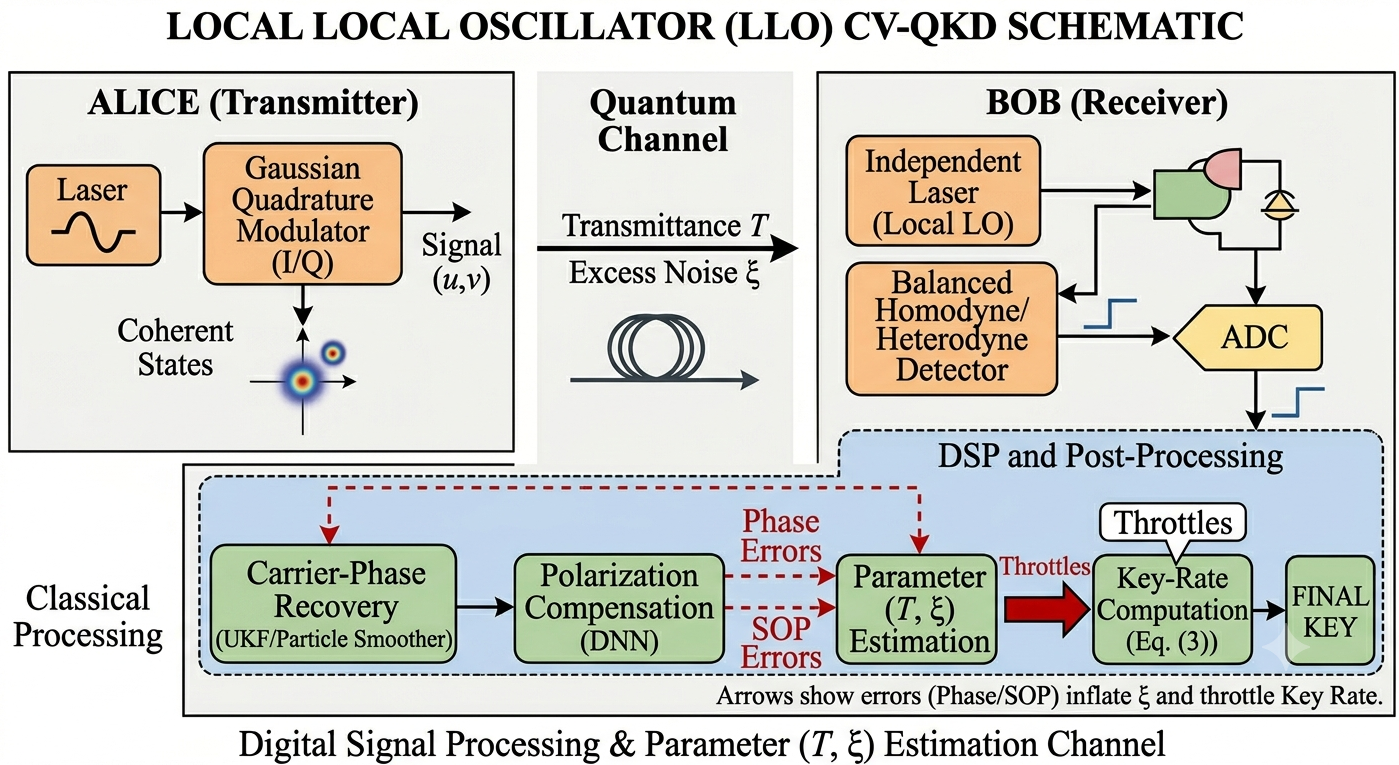}
\caption{Local-local-oscillator CV-QKD, highlighting the digital
signal-processing stage where learned carrier-phase recovery, polarization
compensation and parameter estimation are inserted.}
\label{fig:cvqkd}
\end{figure}

CV-QKD encodes key information in the quadratures of the optical field and
detects with homodyne/heterodyne receivers~\cite{ralph1999cvqkd,
grosshans2002gg02,grosshans2003quantum}. Gaussian-modulated coherent-state
(GG02) protocols are attractive because they use standard telecom
components, but their security and rate depend critically on accurate
estimation of channel transmittance $T$ and excess noise $\xi$. The
asymptotic key rate against collective attacks is
\begin{equation}
R \;=\; \beta\, I_{AB} - \chi_{BE},
\label{eq:cvqkd}
\end{equation}
where $I_{AB}$ is the Alice--Bob mutual information, $\chi_{BE}$ is the
Holevo bound on Eve's information, and $\beta$ is the reconciliation
efficiency~\cite{weedbrook2012gaussian,laudenbach2018cvqkd}. Because Eq.
\eqref{eq:cvqkd} is extremely sensitive to $\xi$, even small uncompensated
phase noise or polarization drift in LLO systems~\cite{qi2015generating,
soh2015selfreferenced} can drive $R\!\to\!0$; learned estimators and filters
address exactly this fragility (Sections~\ref{sec:phase}--\ref{sec:pol}).
Composable finite-size security for CV-QKD is treated
in~\cite{leverrier2010finite,leverrier2015composable}. The corresponding
LLO signal-processing chain is shown in Fig.~\ref{fig:cvqkd}.

\FloatBarrier

\subsection{Security Model, Metrics and Channels}
The practical security of QKD is challenged by device
imperfections~\cite{scarani2009security,xu2020secureqkd} and explicit
quantum-hacking attacks, including detector
blinding~\cite{lydersen2010hacking}, calibration
attacks~\cite{jain2011device}, and CV-specific wavelength and saturation
attacks~\cite{huang2013quantum,qin2016quantum}. Device-independent and
one-sided device-independent (1SDI) formulations~\cite{acin2007device,
branciard2012onesided} trade rate for reduced trust and connect to quantum
steering~\cite{wiseman2007steering,uola2020quantum} (Section~\ref{sec:steer}).
Throughout, we use four recurring performance metrics: \emph{QBER} (DV) and
\emph{excess noise} (CV); \emph{secret key rate} (SKR, bits/s or
bits/pulse); \emph{transmittance/loss}; and \emph{key error rate} (KER).
QKD operates over single-mode fiber, free-space terrestrial links, and
satellite channels; the latter add atmospheric turbulence described by the
Fried parameter and refractive-index structure
constant~\cite{fried1966optical,andrews2005laser,vasylyev2012atmospheric,
vasylyev2016atmospheric}, motivating the channel-characterization learning
of Section~\ref{sec:freespace}.

\subsection{Post-Processing: Reconciliation and Privacy Amplification}
\emph{Information reconciliation} aligns Alice's and Bob's bit strings,
leaking at most $f\!\cdot\!n\!\cdot\!h_2(e)$ bits to Eve; rate-adaptive LDPC
decoders reach $f\!\approx\!1.05$--$1.15$~\cite{laudenbach2018cvqkd}.
\emph{Privacy amplification} then compresses the reconciled string to a key
that is provably independent of Eve's information. Any learned upstream
component (phase estimator, reconciliation decoder) changes only the effective
$e$ and $f$, not the post-amplification security level---provided the induced
rate is computed from the actual $e$ and $f$, not from an optimistic prediction.
For scalable IoT/6G reconciliation, ML-Cascade adaptation is discussed in
Theme~III (Section~\ref{sec:fliot}).

\subsection{Trusted Relay Networks and HAP/6G Coverage}
\begin{figure}[!tb]
\centering
\includegraphics[width=\linewidth]{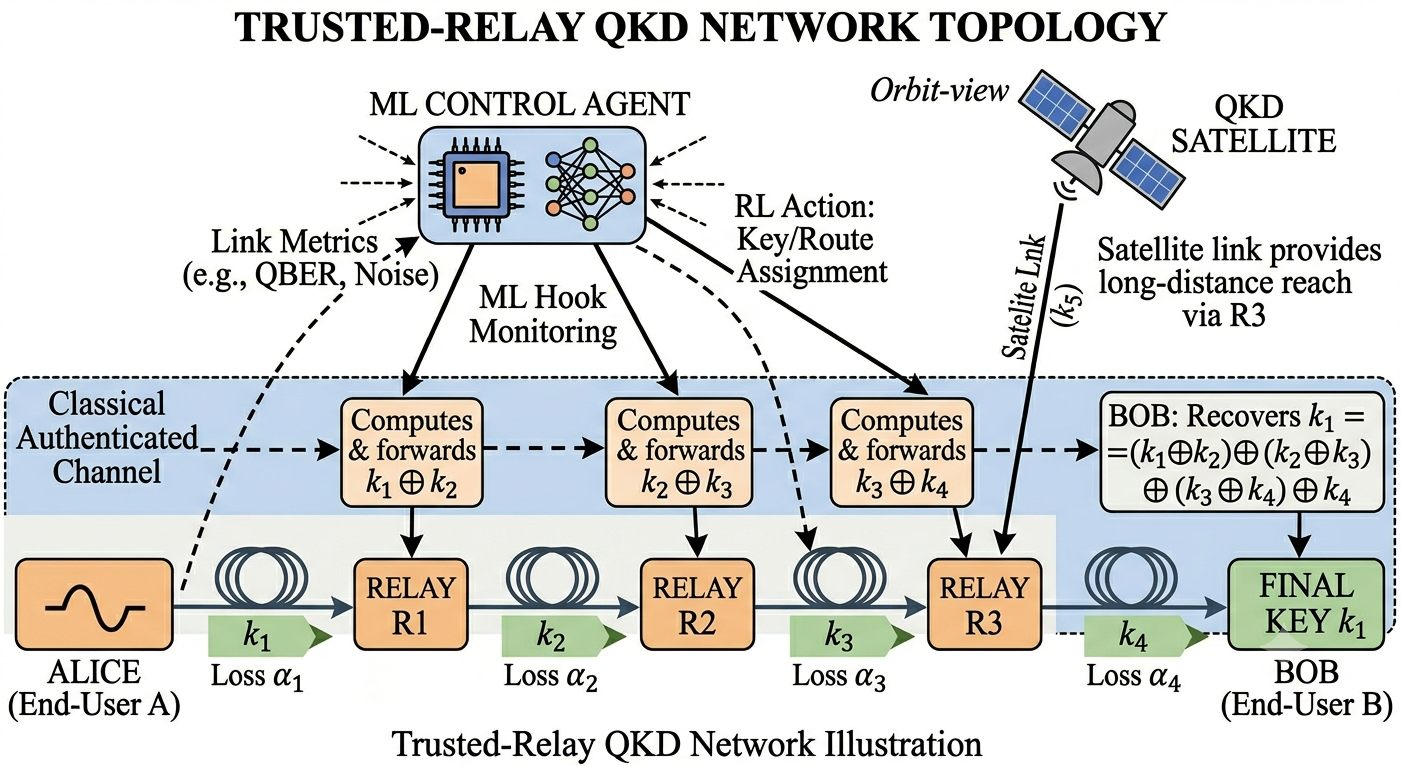}
\caption{Trusted-relay QKD network with XOR key forwarding. ML/RL agents
optimize per-hop rates and network-level key routing for IoT/6G QKD
(Theme~III, Section~\ref{sec:fliot}).}
\label{fig:relay}
\end{figure}

Beyond the repeaterless range, practical QKD uses \emph{trusted relay (TR)}
networks, where each relay decrypts and re-encrypts the key in the classical
domain. Key forwarding uses XOR: if Alice-relay share $k_1$ and relay-Bob
share $k_2$, relay broadcasts $k_1\oplus k_2$ so Bob recovers $k_1$. The
major deployments (SECOQC~\cite{peev2009secoqc}, Tokyo~\cite{sasaki2011field},
the Chinese 4600-km network~\cite{chen2021integrated}) rely on this principle.
For HAP-assisted 6G QKD~\cite{almohammed2025hapxor}, XOR relay composition
carries composable security guarantees through the stratospheric relay chain,
making it directly relevant to Theme~III (Section~\ref{sec:fliot}). The
trusted-relay architecture and XOR forwarding are illustrated in
Fig.~\ref{fig:relay}.

\FloatBarrier

\subsection{Field Trials and Experimental Milestones}
\label{ssec:experiments}
Table~\ref{tab:experiments} summarizes representative experimental QKD
achievements spanning fiber, satellite and integrated networks, as a
reference backdrop for the ML-improvement claims discussed in the survey.

\begin{table*}[!tp]
\centering
\caption{Selected QKD experimental milestones: protocol, medium, distance and
reported key rate. These results provide the performance backdrop against
which the ML-based improvements surveyed in this work are assessed.}
\label{tab:experiments}
\footnotesize
\begin{tabular}{L{2.1cm} c L{1.5cm} L{1.5cm} L{2.0cm} L{2.5cm}}
\toprule
\textbf{Work} & \textbf{Year} & \textbf{Protocol} & \textbf{Medium} & \textbf{Distance} & \textbf{Key rate / notes} \\
\midrule
SECOQC~\cite{peev2009secoqc}                          & 2009 & DV (multi) & Fiber TR  & Metro       & Trusted-relay network, Vienna \\
Tokyo QKD~\cite{sasaki2011field}                      & 2011 & DV (multi) & Fiber TR  & Metro       & 45 km; 304 kbit/s \\
Jouguet \emph{et al.}~\cite{jouguet2013experimental}  & 2013 & CV GG02    & Fiber     & 80 km       & First long-distance CV-QKD \\
MDI 404 km~\cite{yin2016measurement}                  & 2016 & MDI        & Fiber     & 404 km      & Removes detector attacks \\
Satellite-to-ground~\cite{liao2017satellite}          & 2017 & DV BB84    & Satellite & $>$1200 km  & kbit/day \\
Boaron \emph{et al.}~\cite{boaron2018secure}          & 2018 & DV         & Fiber     & 421 km      & $\sim$6.5~bit/s \\
TF ($\sqrt\eta$) proofs~\cite{minder2019experimental,
                               wang2019beating}       & 2019 & TF         & Fiber     & $\sim$200 km & Surpasses repeaterless bound \\
Chen 509 km~\cite{chen2020sending}                    & 2020 & SNS-TF     & Fiber     & 509 km      & $\sim$0.6~bit/s \\
Yin 1120 km~\cite{yin2020entanglement}                & 2020 & E91 (sat.) & Satellite & 1120 km     & Entanglement-based satellite \\
Zhang 202 km CV~\cite{zhang2020long}                  & 2020 & CV GG02    & Fiber     & 202 km      & $\sim$50~bit/s at 200 km \\
Chen 4600 km~\cite{chen2021integrated}                & 2021 & DV+sat. TR & Fiber+Sat & 4600 km     & Space-to-ground network \\
Pittaluga 600 km~\cite{pittaluga2021600}              & 2021 & TF         & Fiber     & 600 km      & Dual-band stabilization \\
Pan 100 km LLO CV~\cite{pan2023cvqkd100km}            & 2024 & CV LLO     & Fiber     & 100 km      & 25.4~kbit/s, 15.4~dB \\
COW 100 km~\cite{experimental_cow_2026}               & 2026 & COW        & Fiber     & 100~km      & kbit/s; finite-key secure \\
\bottomrule
\end{tabular}
\end{table*}
\FloatBarrier

\subsection{Machine Learning Methods}
A concise overview of the ML/RL/QML methods used across the five specialized
themes is provided in Section~\ref{sec:mlmethods}.

\subsection{Finite-Key Effects}
\label{ssec:finitekey}
Asymptotic rates such as Eqs.~\eqref{eq:bb84}--\eqref{eq:cvqkd} are upper
bounds; any real session exchanges a finite number of signals $N$, and the
statistical estimates of yields and error rates carry confidence intervals
that \emph{reduce} the extractable key. A representative finite-key secret
length against general attacks has the structure
\begin{equation}
\ell \;\leq\; n\big[1-h_2(e+\mu_\epsilon)\big] - \mathrm{leak}_{\mathrm{EC}}
   - \log_2\!\tfrac{2}{\epsilon_{\mathrm{cor}}}
   - 2\log_2\!\tfrac{1}{2\epsilon_{\mathrm{PA}}},
\label{eq:finitekey}
\end{equation}
where $n$ is the sifted block length, $\mu_\epsilon$ is a statistical
fluctuation term shrinking as $O(1/\sqrt{n})$, $\mathrm{leak}_{\mathrm{EC}}$
is the information revealed during error correction, and
$\epsilon_{\mathrm{cor}},\epsilon_{\mathrm{PA}}$ are correctness and
privacy-amplification security parameters~\cite{scarani2008quantum,
tomamichel2012tight,leverrier2010finite}. Two consequences matter for ML.
First, the optimal operating parameters depend on $N$, so the parameter
optimization of Section~\ref{sec:paramopt} is genuinely a function of block
size, not just channel loss. Second, because the fluctuation terms are
estimated from data, the quality of \emph{estimation}---of yields, of
excess noise, of phase---feeds directly into the security margin, which is
why so many of the learned components in this survey are estimators.

\subsection{CV-QKD Parameter Estimation}
\label{ssec:cvest}
In CV-QKD the channel is summarized by transmittance $T$ and excess noise
$\xi$, estimated from a disclosed subset of quadrature samples. Eve's
information $\chi_{BE}$ in Eq.~\eqref{eq:cvqkd} is a steep function of $\xi$;
a small estimation bias can either abort an otherwise secure session or, far
worse, overstate secrecy. Classical estimators assume stationarity over the
estimation window, an assumption violated by drifting phase and polarization
in LLO and free-space links. Learned and Bayesian estimators
(Sections~\ref{sec:phase}--\ref{sec:pol}, \ref{sec:freespace}) target
exactly this nonstationarity, tracking $T$ and $\xi$ (or the latent phase and
SOP that corrupt them) as time-varying states rather than fixed unknowns.

\section{ML/RL/QML Methods for Specialized QKD Aspects}
\label{sec:mlmethods}
This section gives a concise account of the model families used in the five
specialized themes. Readers familiar with ML may skip to
Section~\ref{sec:taxonomy}.

\textbf{Tree ensembles (RF, XGBoost, LightGBM)} underpin adaptive protocol
selection and free-space channel prediction. Random forests~\cite{breiman2001random}
average $B$ bootstrap-resampled trees, $\hat y(x)=\frac{1}{B}\sum_b T_b(x)$,
with built-in feature importance; XGBoost~\cite{chen2016xgboost} and
LightGBM~\cite{ke2017lightgbm} are scalable gradient-boosted realizations
for 6G/B5G coexistence and channel-allocation prediction.

\textbf{CNN/LSTM} exploit spatial and temporal structure. CNNs
\cite{krizhevsky2012imagenet,he2016deep} correct OAM phase distortion in
atmospheric QKD and optimize parameters from channel-feature maps. LSTM/GRU
networks~\cite{hochreiter1997lstm,cho2014learning} model temporal dynamics
in HAP/satellite link geometry, keystore depletion history, and time-series
attack patterns.

\textbf{Bayesian estimators}---unscented Kalman filters (UKF)~\cite{julier1997new}
and particle smoothers---track latent states such as carrier phase and excess
noise $\xi$ in adaptive LLO CV-QKD links. Unlike black-box networks, they
carry explicit uncertainty estimates, a key property for security-sensitive
QKD estimators.

\textbf{Reinforcement learning (RL)}~\cite{sutton2018reinforcement} optimizes
sequential decisions via expected discounted return
$J(\pi)=\mathbb{E}_\pi[\sum_t\gamma^t r(s_t,a_t)]$. Deep RL~\cite{mnih2015human}
is applied to IoT/6G key assignment, HAP/UAV resource allocation, and
federated learning support---decisions above the security proof that can be
optimized aggressively.

\textbf{Anomaly detectors}---DBSCAN~\cite{ester1996dbscan} and isolation
forests~\cite{liu2008isolation}---are label-free, which is critical for
IoT/6G attack monitoring where labelled attack data are scarce. Both are
heuristic and evadable, so they function as monitoring layers, never
replacing the security proof.

\textbf{SVM/KNN} are used for steerability classification in 1SDI-QKD
(Theme~V). SVMs~\cite{cortes1995support} find maximum-margin separators
that connect naturally to quantum feature maps in QML~\cite{havlicek2019supervised,
schuld2019quantum}.

\textbf{QML}~\cite{biamonte2017quantum,rebentrost2014quantum,lloyd2013quantum}
embeds data in quantum feature spaces or parameterized quantum circuits.
Specialized uses in this survey include: QLSTM for temporal attack
detection~\cite{alkuwari2026qlstm_iet}; quantum circuit learning for BB84
attack optimization~\cite{decker2025qkd_qml}; and QML-based optimization in
6G networks~\cite{qml_6g}. Whether QML yields a practical advantage for
these tasks remains an open research question (Section~\ref{sec:challenges}).

Table~\ref{tab:mltaxonomy} maps model families to specialized QKD themes.

\begin{table}[!tb]
\centering
\caption{ML/RL/QML model families and their principal use across the five
specialized QKD themes.}
\label{tab:mltaxonomy}
\footnotesize
\begin{tabular}{L{2.5cm} L{5.0cm}}
\toprule
\textbf{Model family} & \textbf{Theme / specialized use} \\
\midrule
RF / boosted trees & I: Adaptive protocol selection; II: FSO/HAP channel prediction \\
SVM / KNN & V: Steerability classification; II: non-terrestrial channel estimation \\
CNN / LSTM & II: OAM correction in atmospheric QKD; I: adaptive parameter support \\
Bayesian filters (UKF) & I: Carrier-phase tracking in adaptive LLO CV-QKD \\
DBSCAN / Isolation Forest & IV: Attack monitoring in IoT/6G QKD; IV: anomaly detection \\
(Deep) RL & III: IoT/6G key assignment; II: HAP/UAV platform control; III: FL support \\
QML / QLSTM & IV: Quantum-enhanced attack detection; V: steerability witnesses \\
\bottomrule
\end{tabular}
\end{table}
\FloatBarrier

\section{Taxonomy: Five Specialized QKD Themes}
\label{sec:taxonomy}
\begin{figure}[!tb]
\centering
\includegraphics[width=\linewidth]{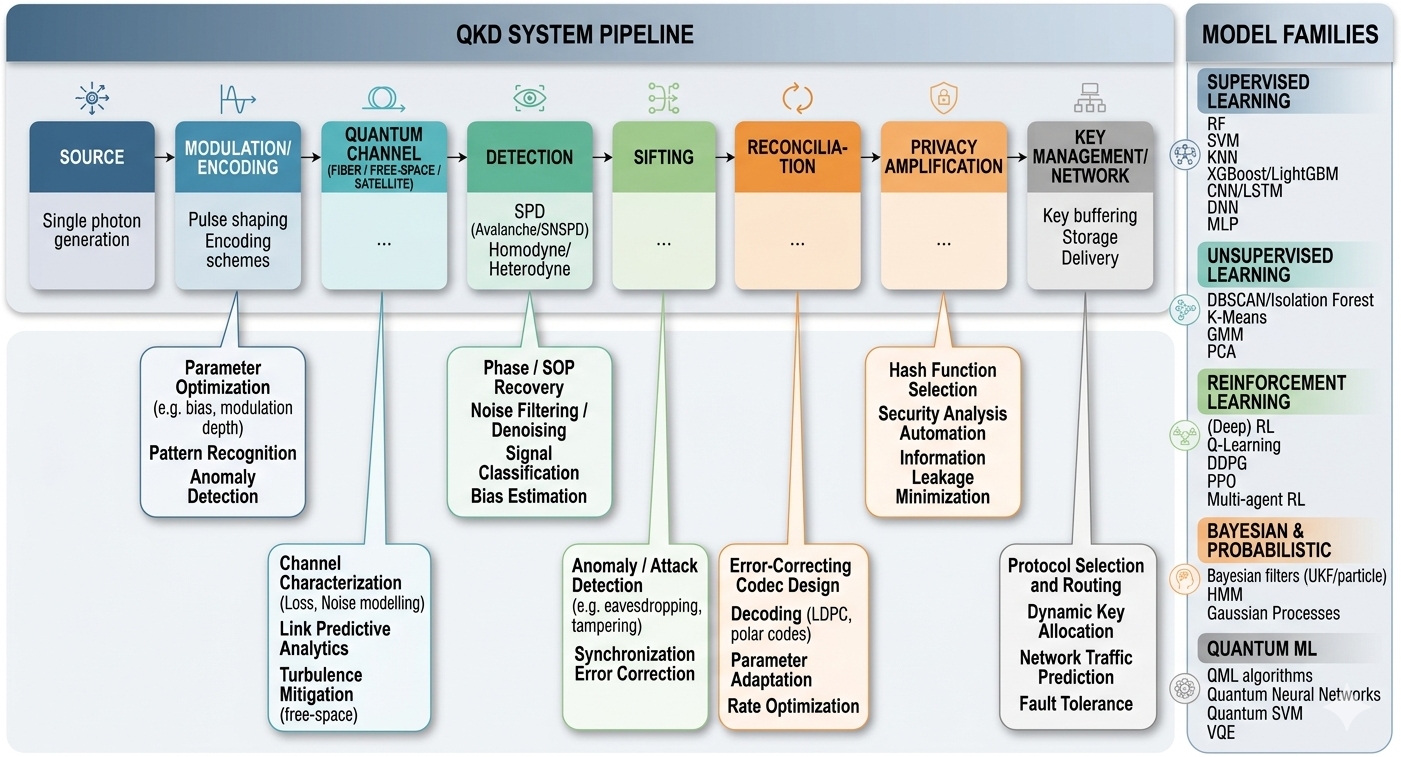}
\caption{Taxonomy mapping ML/RL/QML tasks onto specialized QKD aspects.
The five themes---Adaptive Protocol/Parameter Support, Free-Space/Satellite/UAV/HAP,
IoT/6G/Federated Learning, QML-Assisted Functions, and Steerability/1SDI
Security---define the scope of this survey beyond conventional fiber-link QKD
(Sections~\ref{sec:paramopt}--\ref{sec:steer}).}
\label{fig:taxonomy}
\end{figure}

We surveyed the literature at the intersection of QKD and ML/RL/QML and
organized the works around five specialized themes that reflect the
distinct identity of this survey---going beyond the standard
point-to-point fiber-link pipeline.
These themes are summarized in
Table~\ref{tab:domains} and mapped in
Fig.~\ref{fig:taxonomy}. The themes are defined around
\emph{engineering problems} in specialized QKD deployment scenarios,
because the same algorithm (e.g.\ a random forest) recurs in very different
roles, and because practitioners typically arrive with a scenario
(``my QKD link is over a HAP channel'' or ``which protocol fits this
IoT node?'') rather than with a method.

\FloatBarrier

For every theme, the survey sections present: (i)
the \textbf{problem}, including the physical or networking quantity to be
estimated, controlled or decided; (ii) the \textbf{classical solution}
representing the pre-ML state of the art; (iii) the
\textbf{ML/RL solution}, with the model family, inputs/outputs, training
regime, and reported metrics; and (iv) a short \textbf{analysis} of when
and why learning helps, its costs, and its security caveats, followed by
a per-theme comparison table.

\begin{table}[!tb]
\centering
\caption{The five specialized QKD themes, the deployment scenarios they
address, and their dominant model families.}
\label{tab:domains}
\footnotesize
\begin{tabular}{c L{3.4cm} L{2.4cm}}
\toprule
\textbf{\S} & \textbf{Specialized Theme} & \textbf{Dominant models} \\
\midrule
\ref{sec:paramopt} & Adaptive Protocol \& Parameter Support & NN, RF, RL \\
\ref{sec:freespace}& Free-Space, Satellite, UAV, HAP-Assisted QKD & RF, BPNN, CNN \\
\ref{sec:fliot}    & QKD for IoT, 6G \& Quantum-Secured FL & DRL, QML \\
\ref{sec:qml}      & QML-Assisted QKD Functions             & DBSCAN, QML, QLSTM \\
\ref{sec:steer}    & Steerability-Aware \& 1SDI-QKD Security & SVM, NN \\
\bottomrule
\end{tabular}
\end{table}
\FloatBarrier

Supporting material on adaptive sifting and reconciliation
(Section~\ref{sec:sift}), carrier-phase recovery for adaptive CV-QKD
(Section~\ref{sec:phase}), and polarization tracking for aerial links
(Section~\ref{sec:pol}) is integrated as short supporting subsections
within the five themes above. Network management and coexistence with
6G/B5G classical channels are incorporated as subsections of
Theme~III (Section~\ref{sec:fliot}).

\section{Theme~I: Adaptive Protocol and Parameter Support}
\label{sec:paramopt}
\subsection{Why Adaptive Support Is a Specialized QKD Aspect}
A key challenge in non-terrestrial, mobile, and heterogeneous QKD
deployments is that no single protocol or fixed parameter set is optimal
across all conditions. Free-space links experience time-varying
transmittance, HAP channels have geometry-dependent loss, and IoT
deployments require lightweight protocol decisions. This makes adaptive
protocol selection and real-time parameter support a genuinely specialized
QKD function---different from the static offline optimization assumed in
conventional point-to-point fiber designs. ML is the natural engine for
this adaptation: it can learn the mapping from measured channel/hardware
conditions to optimal protocol and parameter choices, and execute it at
the timescales required by dynamic deployments.

\subsection{Parameter Optimization and SKR Prediction}
\textbf{Problem.} As Eq.~\eqref{eq:decoy} and Eq.~\eqref{eq:cvqkd} show, the
achievable key rate depends on a vector of tunable
parameters---signal/decoy intensities, basis and modulation probabilities,
block lengths, and, in CV-QKD, the modulation variance. The optimal setting
depends on the instantaneous channel loss and noise, which drift in
deployed links (and drift especially rapidly in free-space, satellite, and
mobile channels), so the parameters must be re-optimized repeatedly and,
ideally, in real time.

\textbf{Classical solution.} The textbook approach numerically maximizes the
finite-key rate over the parameter vector for each estimated channel
condition, using local or global optimizers and, for CV-QKD, a convex
minimization over Eve's attacks to bound $\chi_{BE}$~\cite{lo2005decoy,
scarani2008quantum,leverrier2010finite}. This is accurate but
computationally heavy; performing it per-frame on an embedded controller is
often infeasible, and for CV-QKD the key-rate computation alone can dominate
the latency budget.

\textbf{ML/RL solution.} Wang and Lo~\cite{wang2019nnparams} train a neural
network to predict the optimal decoy-state parameters directly from the
channel description for BB84, MDI- and TF-QKD, reproducing the
optimizer's output while reducing inference to a single forward pass---a
speedup of orders of magnitude that makes per-frame re-optimization
practical. For CV-QKD, learned regressors estimate the secret key rate
itself, replacing the convex minimization and reportedly cutting key-rate
computation time by several orders of magnitude with negligible accuracy
loss~\cite{ml_keyrate_estimation,review_cvqkd_ml_2023}. A complementary line
uses CNNs to map measured network/channel features to optimized operating
points, with one study reporting a $\sim$15\% key-rate increase and
$\sim$20\% error reduction relative to a static
configuration~\cite{cnn_qkd_netopt}. The most recent 2026 review of
ML-for-QKD~\cite{review_ml_qkd_2024} reports deep-RL based adaptive
optimization achieving SKR gains of 15--20\% under noisy conditions while
also suppressing QBER by 30--40\%~\cite{shingne2025drl_vae}.
Automated ML with Bayesian optimization has also been applied to discrete-modulation
CV-QKD, finding network architectures that compute key rates with
99.15--99.59\% reliability and a $\sim 10^7$ speedup~\cite{ml_keyrate_estimation}.
OptiQKD~\cite{optiqkd2025} provides a protocol-agnostic unified framework
(TCN + RL) for simultaneous SKR maximization and QBER minimization across
BB84, E91 and COW.

\textbf{Analysis.} Learning here acts as a fast \emph{surrogate} for an
expensive but well-understood optimization. The security argument is
inherited from the underlying rate formula provided the predicted parameters
are treated as a proposal whose induced rate is re-verified; the danger is
using a learned rate \emph{estimate} directly in a security claim, since a
mispredicted $\chi_{BE}$ could overstate secrecy. The practical payoff is
latency: surrogates turn an offline computation into a real-time control
signal. Table~\ref{tab:paramopt} summarizes the domain.

\begin{table}[!tb]
\centering
\caption{Adaptive parameter optimization and SKR prediction: representative
works, models and reported benefits.}
\label{tab:paramopt}
\footnotesize
\begin{tabular}{L{2.0cm} L{2.0cm} L{2.6cm}}
\toprule
\textbf{Work} & \textbf{Model} & \textbf{Reported benefit} \\
\midrule
Wang--Lo~\cite{wang2019nnparams} & Feed-forward NN & Optimal params for BB84/MDI/TF; $\sim 10^{2}$--$10^{3}\times$ faster \\
CV-QKD SKR~\cite{ml_keyrate_estimation} & MLP regressor & Key-rate calc.\ several orders faster \\
CNN net-opt~\cite{cnn_qkd_netopt} & CNN & $+15\%$ key rate, $-20\%$ error \\
\bottomrule
\end{tabular}
\end{table}
\FloatBarrier

\subsection{Adaptive Protocol Selection}
\label{sec:protocol}
\textbf{Problem.} No single QKD protocol is optimal across all conditions.
BB84/decoy excels at short range; MDI removes detector attacks; TF
maximizes reach; CV-QKD suits metro distances with telecom hardware. In
adaptive or mobile deployments---HAPs, trains, UAVs, 6G nodes---channel
conditions change continuously, so the system must select and switch
protocols dynamically rather than applying a fixed offline choice.

\textbf{Classical solution.} Operators apply static rules of thumb or
exhaustively evaluate each candidate's rate formula---accurate but slow and
inflexible as conditions change, and completely unsuitable for real-time
adaptive switching in non-terrestrial scenarios.

\textbf{ML/RL solution.} Ren \emph{et al.}~\cite{ren2021implementation}
train a random forest to select the optimal protocol from channel features,
reporting $>$98\% accuracy. Nayana \emph{et al.}~\cite{nayana2022selector}
build an RF-based selector across BB84/MDI/TF for next-generation networks,
and RF with PCA preprocessing has been used for real-time joint
protocol/resource selection at similar accuracy~\cite{rf_resource_allocation}.
The CV-QKD survey notes the same idea could extend to protocol
\emph{design}, not just selection~\cite{review_cvqkd_ml_2023}.
OptiQKD~\cite{optiqkd2025} provides a protocol-agnostic RL framework
that switches adaptively between BB84, E91 and COW based on measured
channel state, representing the most complete adaptive-switching
demonstration to date.

\textbf{Analysis.} Adaptive protocol selection is a clean supervised
classification problem with abundant simulated training data and a clear
metric (accuracy against the optimizer's choice), which explains the
consistently high reported scores. The chosen protocol's security is
established by its own proof, so the classifier is a low-risk efficiency
layer; the open question is robustness to channel conditions unseen during
training, which is especially important for non-terrestrial scenarios with
large environmental variability. Table~\ref{tab:protocol} summarizes.

\begin{table}[!tb]
\centering
\caption{Adaptive protocol selection: representative works, models, protocol
sets and reported accuracy.}
\label{tab:protocol}
\footnotesize
\begin{tabular}{L{2.0cm} L{2.0cm} L{2.3cm}}
\toprule
\textbf{Work} & \textbf{Model} & \textbf{Protocols / acc.} \\
\midrule
Ren \emph{et al.}~\cite{ren2021implementation} & Random forest & $>$98\% accuracy \\
Nayana \emph{et al.}~\cite{nayana2022selector} & Random forest & BB84/MDI/TF \\
RF+PCA~\cite{rf_resource_allocation} & RF + PCA & $>$98\%, joint resource sel. \\
OptiQKD~\cite{optiqkd2025} & TCN+RL & Adaptive BB84/E91/COW \\
\bottomrule
\end{tabular}
\end{table}
\FloatBarrier

\subsection{Supporting Note: ML-Assisted Sifting and Reconciliation}
\label{sec:sift}
\textbf{Problem.} After detection, raw data contain frames corrupted by
noise, bursts, and---potentially---adversarial manipulation; these must be
sifted out, and the surviving correlated data must be reconciled into an
identical bit string with minimal information leakage. Both steps affect the
final rate via the factors $q$ and $f$ (resp.\ $\beta$) in
Eqs.~\eqref{eq:decoy}--\eqref{eq:cvqkd}.

\textbf{Classical solution.} Sifting traditionally applies fixed statistical
thresholds, while reconciliation uses multi-level coset or LDPC/multi-edge
codes whose decoders are tuned to a target SNR~\cite{laudenbach2018cvqkd}.
Fixed thresholds misfire under nonstationary noise, and decoders optimized
for one operating point lose efficiency away from it.

\textbf{ML/RL solution.} Jin \emph{et al.}~\cite{jin_isoforest} cast
CV-QKD key sifting as anomaly detection: an isolation
forest~\cite{liu2008isolation}, with its split parameters tuned by a genetic
algorithm, isolates anomalous frames that fixed thresholds miss, improving
the quality of the sifted data. For reconciliation, deep-neural-network
decoders have been reported to lower the frame-error rate relative to
linear-fitting and baseline MLP decoders, particularly at higher SNR, while
a complexity analysis highlights faster reconciliation~\cite{ml_keyrate_estimation,
review_cvqkd_ml_2023}.

\textbf{Analysis.} Anomaly-based sifting is attractive because it is
\emph{unsupervised}---it needs no labelled attacks---but it raises the same
caution as any security-relevant learned filter: an adversary aware of the
detector may shape attacks to look ``normal,'' so such filters should
augment, not replace, the security analysis. Learned reconciliation, by
contrast, affects only efficiency $\beta$/$f$, not the security proof, and
is therefore a low-risk, high-value target. Table~\ref{tab:sift} summarizes.

\begin{table}[!tb]
\centering
\caption{ML-assisted sifting and reconciliation: representative works, models
and reported benefits.}
\label{tab:sift}
\footnotesize
\begin{tabular}{L{2.0cm} L{2.2cm} L{2.4cm}}
\toprule
\textbf{Work} & \textbf{Model} & \textbf{Reported benefit} \\
\midrule
Jin \emph{et al.}~\cite{jin_isoforest} & Isolation Forest + GA & Better anomaly removal in sifting \\
DNN decoder~\cite{ml_keyrate_estimation} & Deep NN & Lower frame-error rate at high SNR \\
\bottomrule
\end{tabular}
\end{table}
\FloatBarrier

\subsection{Supporting Note: Adaptive Phase Recovery for LLO CV-QKD}
\label{sec:phase}
\textbf{Problem.} Generating the local oscillator locally
(LLO)~\cite{qi2015generating,soh2015selfreferenced} removes a major CV-QKD
side channel but leaves a fast, fluctuating phase difference between Alice's
and Bob's lasers. Because Eq.~\eqref{eq:cvqkd} is acutely sensitive to
excess noise $\xi$, residual phase error directly throttles the key rate and
can extinguish it.

\textbf{Classical solution.} Pilot-tone schemes interleave reference pulses
and estimate phase by linear (Wiener-type) filtering~\cite{wiener_filter} or
extended Kalman filtering. These work well near design conditions but
degrade under strong phase noise, low pilot SNR, or model mismatch.

\textbf{ML/RL solution.} Hajomer \emph{et al.}~\cite{hajomer2022cvqkd60km}
demonstrated CV-QKD over 60~km of fiber with a real LO using a
machine-learning phase-noise estimator that outperforms conventional
compensation. Pan \emph{et al.}~\cite{pan2023cvqkd100km} extended LLO CV-QKD
to 100~km with an ML-assisted carrier-recovery pipeline, reporting a secret
key rate of 25.4~kbit/s at 15.4~dB channel loss. Bayesian sequential
estimators have also been applied: an unscented Kalman
filter~\cite{ukf_cvqkd_llo,julier1997new} tracks phase/excess-noise jointly,
and a particle smoother with an MCMC step~\cite{particle_smoother_cvqkd}
enables pilot-free operation at very low SNR (reported minimum
$\mathrm{SNR}_b\!\approx\!-6.9$~dB and $5.7\times10^{-4}$~bit/symbol at
26~km). Joint polarization-and-phase learning is treated
in~\cite{chin2022jointpol}. A 2025 neural-network estimator designed
specifically for excess-noise estimation under composable finite-size security
demonstrates that learned estimators can be combined with finite-key security
proofs when designed conservatively~\cite{liu2025nn_excessnoise}.

\textbf{Analysis.} This is the clearest ``ML-as-better-estimator'' story in
QKD: the learned/Bayesian recovery improves an \emph{estimate} that feeds a
standard rate formula, so it raises rate and reach without touching the
security model. The main cost is computational---particle smoothers are
heavier than Wiener filters---so the practical question is the
accuracy/latency trade-off on the receiver's hardware.
Table~\ref{tab:phase} summarizes.

\begin{table}[!tb]
\centering
\caption{Carrier-phase and frequency recovery for LLO CV-QKD: models, link
reach and reported results.}
\label{tab:phase}
\footnotesize
\begin{tabular}{L{1.9cm} L{1.8cm} L{2.6cm}}
\toprule
\textbf{Work} & \textbf{Model} & \textbf{Reported result} \\
\midrule
Hajomer~\cite{hajomer2022cvqkd60km} & ML phase est. & 60~km LLO CV-QKD \\
Pan \emph{et al.}~\cite{pan2023cvqkd100km} & ML carrier rec. & 100~km, 25.4~kbit/s @ 15.4~dB \\
UKF~\cite{ukf_cvqkd_llo} & Unscented KF & Joint phase/noise tracking \\
Particle smoother~\cite{particle_smoother_cvqkd} & Particle + MCMC & Pilot-free, $\mathrm{SNR}_b\!\approx\!-6.9$~dB \\
\bottomrule
\end{tabular}
\end{table}
\FloatBarrier

\subsection{Supporting Note: Polarization Tracking in Aerial and Mobile Links}
\label{sec:pol}
\textbf{Problem.} Fiber birefringence---especially in aerial cable exposed
to wind and temperature---rotates the state of polarization (SOP) on
millisecond timescales, misaligning polarization-encoded qubits and
inflating QBER.

\textbf{Classical solution.} Electronic polarization controllers driven by
gradient or dithering feedback continuously hunt for the alignment that
minimizes error; they track slow drift but can lose lock during fast
excursions and waste key during re-convergence.

\textbf{ML/RL solution.} A lightweight DNN can \emph{predict} the SOP
trajectory and pre-compute the compensating rotation on the Poincar\'e
sphere, so the controller anticipates rather than chases the drift; one
study reports a QBER reduced by a factor of $\sim$3.88, a key-error-rate
improvement of $\sim$89\%, and an SOP-prediction RMSE of 0.007~rad over
aerial fiber~\cite{ml_sop_tracking}. For entanglement-based links, a
stochastic-optimization feedback loop minimizes QBER by searching the
compensator settings~\cite{shi2021fibre}, and joint polarization/phase
estimators serve CV-QKD~\cite{chin2022jointpol}.

\textbf{Analysis.} As with phase recovery, the learned predictor improves an
estimate feeding a standard controller, so security is unaffected and the
benefit is higher availability (less key lost to re-locking). The key risk
is distribution shift: a model trained on one route's wind/temperature
statistics may not transfer, arguing for online adaptation.
Table~\ref{tab:pol} summarizes.

\begin{table}[!tb]
\centering
\caption{SOP tracking and polarization compensation: mechanisms and reported
metrics.}
\label{tab:pol}
\footnotesize
\begin{tabular}{L{1.9cm} L{2.1cm} L{2.3cm}}
\toprule
\textbf{Work} & \textbf{Mechanism} & \textbf{Reported metric} \\
\midrule
Aerial-fiber SOP~\cite{ml_sop_tracking} & DNN predictor + Poincar\'e plan & QBER $/3.88$; KER $+89\%$; RMSE 0.007~rad \\
Shi \emph{et al.}~\cite{shi2021fibre} & Stochastic-opt.\ feedback & Minimized QBER (entanglement link) \\
Chin \emph{et al.}~\cite{chin2022jointpol} & ML joint pol/phase & CV-QKD over installed fiber \\
\bottomrule
\end{tabular}
\end{table}
\FloatBarrier

\section{Theme~II: Free-Space, Satellite, UAV, and HAP-Assisted QKD}
\label{sec:freespace}
\subsection{Why Non-Terrestrial Links Are a Specialized QKD Aspect}
Free-space and non-terrestrial QKD channels---satellite links, UAV
platforms, and high-altitude platforms (HAPs)---are categorically different
from static fiber links. They traverse turbulent atmosphere that fluctuates
the transmittance, introduces beam wander and pointing errors, and imposes
rapidly time-varying geometry. These conditions make standard fiber-QKD
parameter optimization and channel models inapplicable, and demand
specialized learning approaches for channel characterization, link
scheduling, adaptive optics, and platform control.

\subsection{Atmospheric and Free-Space Channel Characterization}
\textbf{Problem.} Free-space and satellite QKD~\cite{liao2017satellite,
yin2020entanglement,bedington2017progress,sidhu2021advances} traverse
turbulent atmosphere that fluctuates the transmittance and distorts the
wavefront, degrading entanglement quality and key rate. Characterizing and
predicting channel quality---and compensating spatial-mode
distortion---under turbulence is the core problem.

\textbf{Classical solution.} Atmospheric models (log-normal/elliptic-beam
transmittance, Kolmogorov turbulence with the Fried
parameter)~\cite{fried1966optical,andrews2005laser,vasylyev2012atmospheric,
vasylyev2016atmospheric,ruppert2019fading} and adaptive-optics hardware
predict and correct distortion but need many measured atmospheric parameters
and costly wavefront sensing.

\textbf{ML/RL solution.} Random-forest regressors predict the Strehl ratio
of a free-space quantum channel with MAPE $\approx$4.44\% (improving to
3.86\% when a fidelity feature is added)~\cite{rf_strehl}, and related RF
models reconstruct channel/density-matrix properties from turbulence
strength~\cite{rf_turbulence}. For orbital-angular-momentum encoding, a CNN
predicts turbulence-induced phase distortion and drives a spatial-light
modulator for adaptive correction~\cite{cnn_oam_turbulence}. A
back-propagation neural network optimizes the modulation variance of
four-state CV-QKD over an elliptic-beam atmospheric channel evaluated by
Monte-Carlo~\cite{bpnn_cvqkd_atmos}, and physically-constrained ML tunes
SPDC sources for higher-dimensional entanglement~\cite{ml_spdc_polarization}.
ML for broader quantum-channel characterization and inverse system design
spans multi-band and space-division-multiplexed systems~\cite{ml_qcc}.

\subsubsection{QKD over FSO links for high-speed transportation}
A practically important and growing sub-domain is QKD secured by free-space
optical (FSO) links aboard high-speed transportation---trains, evacuated
tubes, and UAVs---where atmospheric turbulence, platform vibration and the
Doppler effect create rapidly varying channel conditions.
Al-Mohammed \emph{et al.}~\cite{almohammed2024qkdfso_trains} analyzed the
integration of QKD with FSO for securing communications in high-speed trains
running at several hundred km/h, demonstrating that standard BB84 decoy-state
protocols can achieve positive key rates under a range of weather and link
geometries. An earlier study~\cite{almohammed2023fso_trains} characterized
the FSO channel itself under varying visibility for train scenarios, providing
the channel model used in subsequent ML-based system optimization. An even
more extreme scenario---ultra-high-speed trains in evacuated tubes---was
analyzed in~\cite{almohammed2022fso_tube}, establishing the FSO link budget
under near-vacuum conditions. UAV (unmanned aerial vehicle) platforms
introduce additional challenges of platform attitude fluctuation and varying
altitude; tradeoffs in FSO communications over UAV networks under weather
variation are studied in~\cite{almohammed2024fso_uav}, providing a basis for
ML-assisted link adaptation.

\subsubsection{High-altitude platform (HAP) based QKD}
High-altitude platform stations (HAPS), operating in the stratosphere at
altitudes of 17--22~km, represent an emerging paradigm for quantum-secured
coverage: they combine the large footprint of a satellite with lower latency,
higher link availability (due to smaller elevation angle range), and the
potential for re-pointing~\cite{almohammed2026hap,almohammed2025hapxor}.
Al-Mohammed and Yaacoub~\cite{almohammed2025hapxor} proposed a composable
XOR-relay QKD architecture over HAPs for 6G networks, combining trusted-relay
topology with XOR-based key forwarding and composable security, as an
alternative to direct satellite-to-ground links when repeaterless range is
insufficient. The broader integration of QKD, FSO and HAPs for enhanced IoT
networks is treated in~\cite{almohammed2026hap}. From an ML perspective, HAP
channels present unique conditions: the stratospheric tropospheric boundary,
beam wander, and time-varying link geometry call for predictive channel
models analogous to those developed for satellite links, with the addition of
quasi-static platform dynamics that adaptive RL controllers can exploit
(Section~\ref{sec:fliot}).

\textbf{Analysis.} Transportation, HAP, and satellite QKD sit at the
intersection of free-space channel modeling and adaptive network optimization.
The key ML contributions are: (i) surrogate channel models (RF, BPNN)
replacing slow atmospheric simulations for real-time link scheduling and
parameter adaptation; (ii) CNN adaptive optics for OAM distortion correction;
and (iii) RL-based platform pointing, handover, and link-scheduling controllers
for UAV and HAP platforms. Atmospheric channels are inherently nonstationary
and hard to model from first principles, which is exactly why data-driven
predictors excel here---reported MAPE values in the low single digits are
practically useful for link scheduling. The main obstacle is acquiring
representative training data across weather conditions, platform altitudes,
and link geometries, motivating physics-informed and transfer-learning
approaches for non-terrestrial QKD.
Table~\ref{tab:freespace} summarizes all free-space, satellite, transportation,
and HAP results.

\begin{table}[!tb]
\centering
\caption{Free-space, satellite, transportation and HAP channel
characterization: models, tasks and reported metrics.}
\label{tab:freespace}
\footnotesize
\begin{tabular}{L{1.8cm} L{2.1cm} L{2.5cm}}
\toprule
\textbf{Work} & \textbf{Model} & \textbf{Task / metric} \\
\midrule
Strehl RF~\cite{rf_strehl} & Random forest & Atmospheric MAPE 4.44\% (3.86\%) \\
Turbulence RF~\cite{rf_turbulence} & Random forest & Channel reconstruction \\
OAM CNN~\cite{cnn_oam_turbulence} & CNN + SLM & Adaptive phase correction \\
CV-QKD BPNN~\cite{bpnn_cvqkd_atmos} & BP neural net & Modulation-variance opt. \\
Train FSO~\cite{almohammed2024qkdfso_trains} & Analytic/ML & QKD+FSO, high-speed trains \\
FSO visibility~\cite{almohammed2023fso_trains} & Channel model & FSO vs.\ weather, train link \\
Evacuated tube~\cite{almohammed2022fso_tube} & Link budget & Ultra-high-speed FSO \\
UAV tradeoffs~\cite{almohammed2024fso_uav} & Analysis & FSO-UAV weather tradeoffs \\
HAP QKD~\cite{almohammed2026hap} & System design & QKD+FSO+HAP for IoT \\
HAP XOR~\cite{almohammed2025hapxor} & Protocol+composable & 6G composable relay QKD \\
Free-space opt.~\cite{chandravanshi2025fso_qkd} & Design & Optimizing secure key bits \\
\bottomrule
\end{tabular}
\end{table}
\FloatBarrier

\section{Theme~III: QKD for IoT, 6G, and Quantum-Secured Federated Learning}
\label{sec:fliot}
\subsection{Why IoT/6G Integration Is a Specialized QKD Aspect}
Emerging applications---federated learning across edge devices, massive IoT,
and 6G networks---create specialized QKD requirements that do not arise in
conventional fiber-link deployments: severe resource constraints on IoT
devices, heterogeneous network topologies, high mobility, dynamic key demand,
and the dual role of ML as both a workload to secure and a tool to optimize
key distribution. QKD in this context is not simply a longer or faster fiber
link---it is a fundamentally different deployment paradigm that requires
learning-assisted orchestration at every level.

\textbf{Problem.} Emerging applications need scalable key distribution and
privacy guarantees, while themselves generating the ML workloads that QKD
must secure.

\textbf{Classical solution.} Classical key management and post-quantum
cryptography secure these systems today but lack information-theoretic
guarantees; static resource allocation underutilizes scarce quantum
resources.

\textbf{ML/RL solution.} Across IoT, 6G, and quantum-secured federated-learning
scenarios, resource allocation, key provisioning, and key-demand prediction
require specialized ML/RL approaches. Federated learning has been
integrated with quantum computing for privacy-preserving distributed
models~\cite{fl_quantum}; a metaverse and 6G synergy survey identifies
QKD as a key enabling layer for semantic communications and edge
learning~\cite{aloudat2025metaverse}. Drone-mounted mobile QKD has been
proposed to secure IoT devices, with weather-aware design
guidelines~\cite{drone_qkd_iot}; and QML has been explored for real-time
optimization and resource allocation in beyond-5G/6G
networks~\cite{qml_6g,nawaz2019quantum,wang2021prospect}. Earlier work
established the architectural principles for integrating quantum
communications with IoT in the 6G era~\cite{almohammed2021icc}.

\subsubsection{ML-augmented cascade reconciliation for scalable QKD}
A key scalability bottleneck is error reconciliation efficiency as network
size grows. Al-Mohammed \emph{et al.}~\cite{almohammed2024cascadeqkd}
propose integrating ML techniques with the \emph{Cascade} protocol: an
autoencoder predicts the QBER of the current block, and its prediction drives
the Cascade block-size selection in real time. This reduces the information
leakage penalty during reconciliation under varying channel conditions, a
task that static block-size policies handle poorly.

\subsubsection{HAP-based QKD for 6G coverage}
A composable XOR-relay QKD architecture over HAPs~\cite{almohammed2025hapxor}
offers 6G-scale coverage with provable composable security, combining the
large-area reach of stratospheric platforms with the information-theoretic
guarantees of QKD, and is complemented by a system-level integration of
QKD, FSO and HAP for IoT~\cite{almohammed2026hap}. These
architectures require ML-based link adaptation (Section~\ref{sec:freespace})
and RL-based resource allocation (Section~\ref{sec:network}) to be viable
at scale. Quantum Radar studies~\cite{almohammed2020icenco_radar} and quantum
computer architecture investigations~\cite{almohammed2020icenco_arch}
provide foundational quantum-hardware context for this convergence.

\subsubsection{Recent system-level advances (2025--2026)}
Deep RL-driven key provisioning for 6G IoT~\cite{seok2025drl_keyprovision}
uses a graph-attention network combined with LSTM to model network topology
and temporal dependencies; it reports significant improvement in session key
availability and reduced keystore exhaustion over greedy baselines---a
critical operational metric for 6G networks with dense IoT traffic. QNN-QRL
frameworks~\cite{behera2025qnnqrl} combine QNN architectures with quantum RL
to improve BB84 and B92 key generation under noisy quantum channels,
evaluating performance with accuracy, F1, and ROC metrics. OptiQKD
\cite{optiqkd2025} proposes a protocol-agnostic ML framework using temporal
convolutional networks plus protocol-aware RL for BB84, E91 and COW.
Shingne \emph{et al.}~\cite{shingne2025drl_vae} report a 15--20\% SKR
gain and 30--40\% QBER reduction through DRL+VAE adaptive QKD optimization
under noisy conditions.

\begin{figure}[!tb]
\centering
\includegraphics[width=\linewidth]{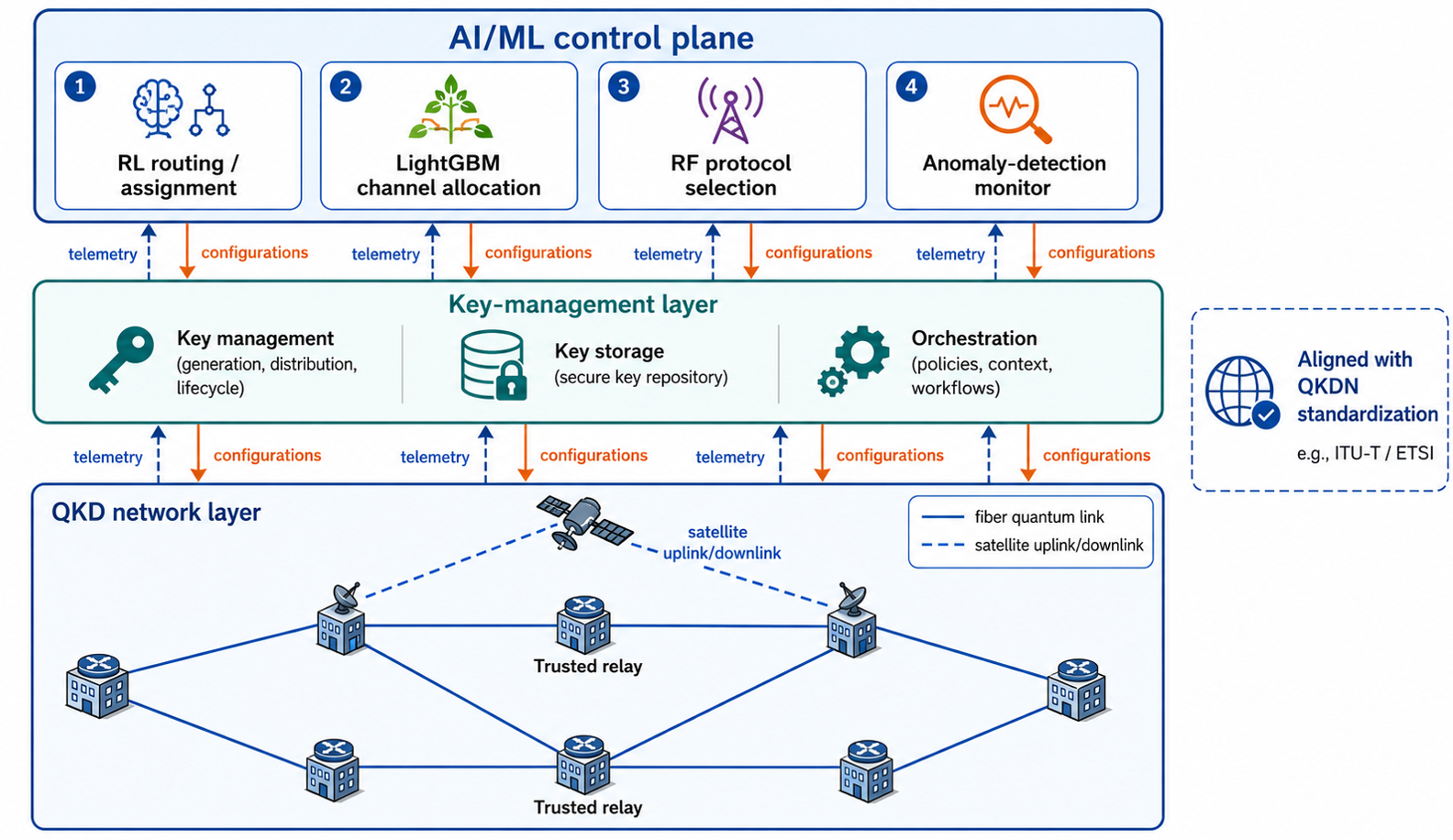}
\caption{Network-level view: an AI/ML control plane consuming link telemetry
and orchestrating routing, allocation, protocol choice and monitoring across
a hybrid fiber/satellite/HAP QKD network (Theme~III, Section~\ref{sec:fliot}).}
\label{fig:network}
\end{figure}

\textbf{Analysis.} This theme is the most application-facing and the least
standardized; ML/RL acts as the orchestration intelligence that makes
quantum security usable at IoT/6G scale. DRL key provisioning
reduces keystore exhaustion, lightweight protocol selection enables
resource-constrained IoT nodes, and HAP/FSO integration extends 6G
coverage. The risk is hype: many results are simulation-only, and the
quantum advantage of native QML in these settings remains to be
demonstrated on hardware. Tables~\ref{tab:network} and \ref{tab:fliot}
summarize the key results, while Fig.~\ref{fig:network} shows the
corresponding AI/ML control-plane view of a hybrid QKD network.

\FloatBarrier

\subsection{DRL for Key Assignment, Routing, and Resource Allocation in IoT/6G}
\label{sec:network}
Dynamic key assignment, routing, and resource allocation under changing
IoT/6G topology require scalable ML. The key-assignment problem on a
QKD-network graph $G=(V,E)$ with link capacities $R_{ij}$ and user demands
$d_k$ can be written as:
\begin{equation}
\max_{\{f_k\}} \sum_k u_k(f_k) \;\text{ s.t. }\; \sum_{k:\, (i,j)\in P_k} f_k \leq R_{ij}
\;\forall (i,j)\in E.
\label{eq:routing}
\end{equation}
RL/DRL addresses this scalably: a deep-RL framework learns an online
key-assignment policy for the one-to-many twinning problem~\cite{rl_onmtp};
deep RL with graph attention and LSTM provisioning~\cite{seok2025drl_keyprovision}
reduces keystore exhaustion in dense IoT networks; multi-agent DRL allocates
resources for quantum-secured federated edge learning~\cite{drl_fel}; and a
two-stage stochastic program for QKD-enabled FL reduces deployment cost by
7.72\%~\cite{qkdfl_stochastic}. Network decisions sit \emph{above} the security
proof, so these optimizations do not affect QKD security guarantees.

\begin{table}[!tb]
\centering
\caption{DRL/RL for IoT/6G QKD key assignment, routing and resource
allocation: models, decision types and reported benefits.}
\label{tab:network}
\footnotesize
\begin{tabular}{L{1.9cm} L{2.1cm} L{2.3cm}}
\toprule
\textbf{Work} & \textbf{Model} & \textbf{Decision / benefit} \\
\midrule
On-MTP~\cite{rl_onmtp} & RL policy & Automated key assignment \\
FEL alloc.~\cite{drl_fel} & Multi-agent DRL & Resource allocation \\
QKD-FL~\cite{qkdfl_stochastic} & Two-stage SP & $-7.72\%$ deployment cost \\
ML-NSCA~\cite{mlnsca} & LightGBM & $\sim$95\% optimal channel \\
DRL provisioning~\cite{seok2025drl_keyprovision} & GAT+LSTM RL & $\downarrow$ keystore exhaustion \\
\bottomrule
\end{tabular}
\end{table}
\FloatBarrier

\subsection{QKD Coexistence with 6G/B5G Classical Channels}
\label{sec:coexist}
In 6G/B5G fronthaul, QKD must coexist with DWDM/SWDM classical traffic
over shared fiber. Spontaneous Raman scattering contaminates the quantum
signal, and analytic models are slow. ML surrogates predict interference
rapidly: supervised regressors (RF, Lasso, Ridge, KNN) integrated with SDN
controllers predict noise, SKR, and QBER directly from measurable channel
parameters~\cite{ml_sdn_coexistence}. For SWDM B5G fronthaul, XGBoost and
LightGBM predict interference with $\sim$98.8\% reduction in computation
time and near-100\% adjacent-channel-power-ratio
accuracy~\cite{swdm_b5g_ml}---enabling rapid channel-plan evaluation for
dynamic 6G spectrum management. Table~\ref{tab:coexist} summarizes the
models, predicted quantities, and reported computational benefits.

\begin{table}[!tb]
\centering
\caption{QKD coexistence with 6G/B5G classical channels: models, predicted
quantities and reported benefits.}
\label{tab:coexist}
\footnotesize
\begin{tabular}{L{1.9cm} L{2.2cm} L{2.2cm}}
\toprule
\textbf{Work} & \textbf{Model} & \textbf{Benefit} \\
\midrule
SDN coexist.~\cite{ml_sdn_coexistence} & RF/Lasso/Ridge/KNN & Noise/SKR/QBER prediction \\
SWDM B5G~\cite{swdm_b5g_ml} & XGBoost/LightGBM & $-98.8\%$ time; $\sim$100\% ACPR \\
\bottomrule
\end{tabular}
\end{table}
\FloatBarrier

\begin{table}[!tb]
\centering
\caption{Quantum-secured federated learning, IoT, HAP and 6G: approaches and
reported metrics.}
\label{tab:fliot}
\footnotesize
\begin{tabular}{L{1.9cm} L{2.1cm} L{2.3cm}}
\toprule
\textbf{Work} & \textbf{Approach} & \textbf{Focus / metric} \\
\midrule
FL+quantum~\cite{fl_quantum} & FL + QC & Privacy-preserving models \\
Drone QKD~\cite{drone_qkd_iot} & Mobile QKD & IoT security, weather-aware \\
IoT 6G~\cite{almohammed2021icc} & System arch. & QComm for IoT in 6G era \\
Cascade ML~\cite{almohammed2024cascadeqkd} & Autoencoder & Scalable reconciliation \\
HAP XOR~\cite{almohammed2025hapxor} & Composable relay & 6G QKD coverage \\
HAP+FSO~\cite{almohammed2026hap} & System integration & QKD+FSO+HAP for IoT \\
Metaverse+6G~\cite{aloudat2025metaverse} & Survey & 6G/edge/semantic comm. \\
DRL key prov.~\cite{seok2025drl_keyprovision} & GAT + LSTM RL & Key availability $\uparrow$ \\
QNN-QRL~\cite{behera2025qnnqrl} & QNN + QRL & BB84/B92 under noisy channel \\
DRL+VAE~\cite{shingne2025drl_vae} & DRL + VAE & SKR $+15$--$20\%$; QBER $-30$--$40\%$ \\
QML 6G~\cite{qml_6g} & QML / QGA & Real-time optimization \\
\bottomrule
\end{tabular}
\end{table}
\FloatBarrier

\section{Theme~IV: QML-Assisted QKD Functions}
\label{sec:qml}
\subsection{QML as the Central Specialized Capability}
This theme is about \emph{quantum} machine learning applied to QKD
functions---a direction categorically different from the classical ML
covered in Themes~I--III. QML embeds QKD data in quantum feature spaces or
processes it with parameterized quantum circuits, potentially probing quantum
channel structure in ways classical networks cannot. QML is not yet a proven
replacement for classical ML in QKD: no current work demonstrates a
practical quantum advantage on real hardware at QKD-relevant scale. But the
emerging results are distinctive enough to deserve a dedicated specialized
theme.

\subsection{Classical ML for Attack Detection: Background Context}
\textbf{Problem.} Quantum-hacking attacks---detector
blinding~\cite{lydersen2010hacking}, calibration
attacks~\cite{jain2011device}, and CV-specific wavelength/saturation
attacks~\cite{huang2013quantum,qin2016quantum}---perturb statistics subtly.
Hardware countermeasures target only \emph{known} attack types. ML provides
attack-agnostic detection, especially in IoT/B5G networks where classical
countermeasures are too heavy: a DBSCAN-based system flags known and unknown
CV-QKD attacks as statistical outliers~\cite{dads_dbscan}. ANN and deep
learning models detect MITM and intercept-resend attacks in B5G IoT QKD links
with $\sim$99\% accuracy, with demonstrated application to high-speed railway
scenarios~\cite{almohammed2021access_ml,almohammed2021gcwkshps}. An ML
study of key-length effects identifies security regimes across
protocols~\cite{almohammed2024bookkeylen}. These classical ML results provide
the baseline against which QML methods below are compared.

\subsection{QML-Enhanced Attack Detection}
\textbf{QLSTM for temporal attack classification.}
Al-Kuwari \emph{et al.}~\cite{alkuwari2025qlstm,alkuwari2026qlstm_iet}
propose a hybrid Quantum Long Short-Term Memory (QLSTM) model that embeds the
QBER and loss-rate time series in quantum feature space. The IET version
reports $\sim$93.7\% accuracy across five attack types
(Intercept-and-Resend, PNS, Trojan-Horse, detector-blinding, and calibration
attacks), outperforming classical LSTM and CNN baselines. The quantum feature
embedding captures temporal correlations in the attack signature that
classical recurrent networks approximate but cannot access
structure-efficiently.

\textbf{QCL for BB84 attack optimization.}
Decker \emph{et al.}~\cite{decker2025qkd_qml} take the complementary
view: framing QKD attack \emph{optimization} as a QML task. Using quantum
circuit learning (QCL), they find the optimal individual attack on BB84---the
phase-covariant cloning machine---without prior knowledge of the analytic
solution. This demonstrates that QML can explore the quantum-attack space
natively, a capability with no direct classical analog.

\subsection{QML for Protocol Support and Optimization}
QML approaches have also been explored for QKD protocol optimization and
resource allocation in 6G networks~\cite{qml_6g,behera2025qnnqrl}. QNN-QRL
frameworks combine quantum neural networks with quantum RL to improve BB84
and B92 key generation under noisy channel conditions~\cite{behera2025qnnqrl}.
These results suggest that QML may offer specialized advantages for
\emph{quantum-native} QKD optimization tasks where the channel model itself
has quantum structure---but the advantage has not yet been demonstrated
against well-tuned classical RL on equal hardware.

\subsection{Realistic Limits of QML for QKD}
Current QML results for QKD share important limitations: (i) all evaluations
are on classical simulators or small quantum devices, not production QKD
hardware; (ii) no published work demonstrates a clear quantum \emph{speedup}
or \emph{accuracy advantage} over classical ML under equal resource conditions;
(iii) the additional overhead of quantum circuit transpilation and
noise-mitigation on near-term hardware may outweigh any intrinsic advantage.
The soundest current posture is to treat QML as an emerging monitoring and
optimization layer---a promising research direction rather than a deployed
tool. Classical ML attack detectors remain more mature and should not be
replaced by QML until a hardware-validated advantage is demonstrated.

\textbf{Analysis.} QML-assisted QKD functions are valuable precisely
because they probe quantum channel structure natively. Classical ML baselines
exist and should be compared fairly. The soundest posture treats all ML/QML
detectors as monitoring layers \emph{above} the security proof, never as
proof elements. Table~\ref{tab:attack} summarizes the key works in this theme.

\begin{table}[!tb]
\centering
\caption{Classical and QML-assisted QKD monitoring and attack-analysis
functions: models, scope and reported metrics.}
\label{tab:attack}
\footnotesize
\begin{tabular}{L{2.0cm} L{2.0cm} L{2.4cm}}
\toprule
\textbf{Work} & \textbf{Model} & \textbf{Scope / metric} \\
\midrule
DADS~\cite{dads_dbscan} & DBSCAN & Known + unknown CV-QKD attacks \\
Imperfection ID~\cite{review_attacks_ml_2024} & Supervised ML & Attack/imperfection classification \\
IoT ANN/DL~\cite{almohammed2021access_ml} & ANN + deep learning & MITM/intercept in B5G IoT, railway; 99\% acc. \\
GC Wkshps~\cite{almohammed2021gcwkshps} & Neural network & IoT QKD attacker detection baseline \\
Key length~\cite{almohammed2024bookkeylen} & ML classifier & Key-length security regime identification \\
QLSTM~\cite{alkuwari2026qlstm_iet} & Hybrid QLSTM & 93.7\% acc.\ over 5 attack types \\
QCL~\cite{decker2025qkd_qml} & QML circuit learning & Optimal BB84 individual attack recovered \\
DRL+VAE~\cite{shingne2025drl_vae} & DRL + VAE & 15--20\% SKR gain, 30--40\% QBER reduction \\
\bottomrule
\end{tabular}
\end{table}
\FloatBarrier

\section{Theme~V: Steerability-Aware and 1SDI-QKD Security Estimation}
\label{sec:steer}
\subsection{Why Steerability Estimation Is a Specialized QKD Aspect}
One-sided device-independent (1SDI) QKD~\cite{branciard2012onesided} bases
its security on EPR steering~\cite{einstein1935can,wiseman2007steering,
uola2020quantum}---a form of quantum nonclassical correlation weaker than
full Bell nonlocality but stronger than entanglement alone. In 1SDI-QKD one
party's device may be entirely untrusted; the protocol's security is
conditional on demonstrating that the shared state is steerable. This
setting is especially relevant for non-ideal channels (atmospheric, HAP,
IoT) where device characterization is incomplete. The central computational
challenge is that deciding steerability and quantifying the steerable
weight are expensive, making real-time online assessment infeasible with
classical SDP alone. ML provides the fast surrogate needed for practical
deployment, making this one of the most distinctive specialized themes in
this survey.

\textbf{Problem.} Deciding whether a given (possibly noisy) bipartite state
is steerable, and quantifying how steerable it is, are computationally
demanding tasks central to 1SDI-QKD security assessment.

\textbf{Classical solution.} Steerability is certified via
semidefinite-programming (SDP) hierarchies and steering
inequalities~\cite{cavalcanti2017quantum}. SDP gives rigorous labels but is
expensive: evaluating it over many states or in real time is impractical for
deployed QKD links.

\textbf{ML/RL solution.} Support-vector-machine classifiers trained on
SDP-labelled data classify states as steerable/unsteerable, and feed-forward
neural networks regress the steerable weight, with a reported accuracy of
$\sim$0.96~\cite{svm_steerability,nn_steerable_weight}. Once trained, these
models replace the SDP with a fast inference suitable for online use.
Both approaches exploit the smooth dependence of steerability on the density
matrix parameters---an ideal regime for kernel methods and shallow neural
networks.

\subsubsection{Physical ML for entanglement source optimization}
The steerability of the distributed two-qubit state also depends on the
entanglement quality of the source, which can be improved by ML tuning of
SPDC source parameters~\cite{ml_spdc_polarization}: physically-constrained
ML modulates the nonlinear crystal's phase-matching conditions to maximize
the visibility of two-photon interference and, consequently, the
entanglement fidelity and steerable weight available to the 1SDI-QKD
session.

\textbf{Analysis.} SVM/NN models act as fast surrogates for the SDP,
enabling online steerability monitoring that would be infeasible with
classical SDP alone. Accuracy must be validated against the SDP ground
truth. Critically, any ML-derived steerable weight used \emph{inside} a
security proof must be a conservative lower bound---an overoptimistic
classifier can overstate the degree of steering and, consequently, overstate
the 1SDI-QKD security level. The Tier~II risk classification
(Section~\ref{ssec:risk}) applies: a pessimistic (lower-bound) estimate of
steerable weight is the only safe use within a security argument. QML
steerability witnesses---using quantum feature maps to probe the density
matrix directly---are a natural next step that would link Theme~IV and
Theme~V, and remain an open research direction~\cite{biamonte2017quantum}.
Table~\ref{tab:steer} summarizes.

\begin{table}[!tb]
\centering
\caption{Steerability estimation for 1SDI-QKD: models, outputs and reported
accuracy.}
\label{tab:steer}
\footnotesize
\begin{tabular}{L{2.0cm} L{2.0cm} L{2.2cm}}
\toprule
\textbf{Work} & \textbf{Model} & \textbf{Output / acc.} \\
\midrule
SVM steer.~\cite{svm_steerability} & SVM (SDP-labelled) & Steerable / not \\
NN weight~\cite{nn_steerable_weight} & Feed-forward NN & Steerable weight, $\sim$0.96 \\
SPDC ML~\cite{ml_spdc_polarization} & Physics-constrained & Source entanglement opt. \\
\bottomrule
\end{tabular}
\end{table}
\FloatBarrier

\section{Consolidated Comparison Across Specialized Themes}
\label{sec:consolidated}

\subsection{Master Reference Table of ML/RL Applications in Specialized QKD}
Table~\ref{tab:master} provides a comprehensive, single-glance reference of
all the primary ML/RL applications surveyed, organized by specialized theme.
It is intended as a look-up table for practitioners entering from a specific
non-terrestrial or application-driven QKD scenario: each row gives the work,
theme, model type, primary input features, optimization target, dataset
source, and headline result, enabling rapid identification of the most
relevant prior art. Table~\ref{tab:master_supp} lists the supporting
functions that provide context for these themes.

\begin{table*}[!tp]
\centering
\caption{Primary ML/RL/QML applications across the five specialized QKD
themes. Themes are: (I) Adaptive Protocol and Parameter Support;
(II) Free-Space, Satellite, UAV and HAP-Assisted QKD; (III) IoT, 6G and
Quantum-Secured Federated Learning; (IV) QML-Assisted QKD Functions; and
(V) Steerability-Aware and 1SDI-QKD Security. ``Sim.'' denotes
simulation-generated data and ``Exp.'' denotes experimental data.}
\label{tab:master}
\scriptsize
\renewcommand{\arraystretch}{1.15}
\begin{tabular}{L{2.3cm} L{1.2cm} L{1.3cm} L{1.6cm} L{2.0cm} c L{2.5cm}}
\toprule
\textbf{Work (year)} & \textbf{Theme} & \textbf{Model} & \textbf{Key inputs} & \textbf{Target} & \textbf{Data} & \textbf{Headline result} \\
\midrule
Wang--Lo '19~\cite{wang2019nnparams}               & I   & Feed-fwd NN   & Channel loss, $\mu, \nu$ & Optimal $(\mu,\nu,p)$ & Sim. & $10^{2}$--$10^{3}\times$ faster; near-optimal \\
CV-KR NN~\cite{ml_keyrate_estimation}              & I   & MLP           & $T,\xi$, block size     & SKR               & Sim. & $\sim 10^7\times$ speedup; 99.15--99.59\% \\
CNN net-opt~\cite{cnn_qkd_netopt}                  & I   & CNN           & Network features        & Operating point   & Sim. & $+15\%$ SKR, $-20\%$ QBER \\
DRL+VAE '25~\cite{shingne2025drl_vae}              & I   & DRL+VAE       & Channel metrics         & SKR, QBER         & Sim. & $+15$--$20\%$ SKR; $-30$--$40\%$ QBER \\
OptiQKD '26~\cite{optiqkd2025}                     & I   & TCN+RL        & Channel state, protocol & SKR, QBER         & Sim. & Protocol-agnostic BB84/E91/COW \\
Ren '21~\cite{ren2021implementation}               & I   & Random forest & Channel features        & Protocol label    & Sim. & $>$98\% accuracy \\
Nayana '22~\cite{nayana2022selector}               & I   & Random forest & Link features           & BB84/MDI/TF       & Sim. & High accuracy, next-gen networks \\
RF+PCA~\cite{rf_resource_allocation}               & I   & RF+PCA        & Network state           & Protocol+resources& Sim. & $>$98\% joint selection \\
Strehl RF~\cite{rf_strehl}                         & II  & Random forest & Atmospheric params      & Strehl ratio      & Sim. & MAPE 3.86\%--4.44\% \\
OAM CNN~\cite{cnn_oam_turbulence}                  & II  & CNN+SLM       & Pupil-plane image       & Phase distortion  & Sim. & Adaptive OAM correction \\
Train FSO~\cite{almohammed2024qkdfso_trains}       & II  & Analysis      & Weather, geometry       & Key rate          & Sim./Exp. & QKD+FSO train feasibility \\
HAP XOR~\cite{almohammed2025hapxor}                & II  & Composable    & HAP link geometry       & Secure relay key  & Analytic & 6G composable HAP QKD \\
HAP+FSO~\cite{almohammed2026hap}                   & II  & System        & QKD+FSO params          & Coverage          & Sim. & QKD+FSO+HAP integration \\
On-MTP RL~\cite{rl_onmtp}                          & III & RL policy     & Network state, demands  & Key assignment    & Sim. & Online policy learning \\
DRL key prov.~\cite{seok2025drl_keyprovision}      & III & GAT+LSTM RL   & Topology, keystore      & Key availability  & Sim. & $\downarrow$ session interruptions \\
QKD-FL~\cite{qkdfl_stochastic}                     & III & Two-stage SP  & Topology, traffic       & Cost              & Sim. & $-7.72\%$ deployment cost \\
ML-NSCA~\cite{mlnsca}                              & III & LightGBM      & Channel features        & Alloc.\ optimum   & Sim. & $\sim$95\% accuracy \\
SDN coexist.~\cite{ml_sdn_coexistence}             & III & RF/Lasso/KNN  & DWDM configuration      & QBER, SKR, noise  & Sim. & Joint SKR/QBER prediction \\
SWDM B5G~\cite{swdm_b5g_ml}                        & III & XGB/LGBM      & SWDM channel params     & Noise             & Sim. & $-98.8\%$ time; $\sim$100\% ACPR \\
DRL FEL~\cite{drl_fel}                             & III & Multi-agent   & Network, channel state  & Resource alloc.   & Sim. & Federated RL allocation \\
QNN-QRL~\cite{behera2025qnnqrl}                    & III & QNN+QRL       & Noisy channel           & BB84/B92 key rate & Sim. & QML under noise \\
DADS~\cite{dads_dbscan}                            & IV  & DBSCAN        & Statistical features    & Attack flag       & Sim. & Known + unknown attacks \\
IoT ANN~\cite{almohammed2021access_ml}             & IV  & ANN+DL        & QBER, timing, counts    & Attacker present? & Sim. & 99\% accuracy, B5G IoT \\
GC Wkshps~\cite{almohammed2021gcwkshps}            & IV  & NN            & QKD statistics          & Attacker present? & Sim. & Baseline IoT attacker detector \\
QLSTM '26~\cite{alkuwari2026qlstm_iet}             & IV  & Hybrid QLSTM  & QBER, loss, time series & 5-class attack    & Sim. & 93.7\% acc., 5 attack types \\
QCL~\cite{decker2025qkd_qml}                       & IV  & QML circuit   & Circuit parameters      & Optimal attack    & Sim. & Recovers PCCM on BB84 \\
Key length ML~\cite{almohammed2024bookkeylen}      & IV  & ML classifier & Key length, protocol    & Security regime   & Sim. & Key-length security analysis \\
SVM steer.~\cite{svm_steerability}                 & V   & SVM+SDP       & Bloch-sphere params     & Steerable?        & Sim. & Binary classification \\
NN weight~\cite{nn_steerable_weight}               & V   & Feed-fwd NN   & State parameters        & Steerable weight  & Sim. & $\sim$0.96 accuracy \\
\bottomrule
\end{tabular}
\end{table*}
\FloatBarrier

\begin{table*}[!tp]
\centering
\caption{Supporting ML functions: sifting, phase recovery, polarization
tracking, network management and coexistence. Rows labelled ``Supp.'' provide
context for the five specialized themes without being primary specialized
contributions of this survey.}
\label{tab:master_supp}
\scriptsize
\renewcommand{\arraystretch}{1.15}
\begin{tabular}{L{2.3cm} L{1.2cm} L{1.3cm} L{1.6cm} L{2.0cm} c L{2.5cm}}
\toprule
\textbf{Work (year)} & \textbf{Theme} & \textbf{Model} & \textbf{Key inputs} & \textbf{Target} & \textbf{Data} & \textbf{Headline result} \\
\midrule
Jin IF~\cite{jin_isoforest}                        & Supp. & Iso.\ Forest  & Raw quadrature samples  & Anomaly flag      & Sim. & Improved sifting quality \\
DNN decoder~\cite{ml_keyrate_estimation}           & Supp. & Deep NN       & Syndrome, frame data    & Error correction  & Sim. & Lower FER at high SNR \\
Hajomer '22~\cite{hajomer2022cvqkd60km}            & Supp. & ML est.       & Pilot+data quadratures  & Phase noise       & Exp. & 60~km LLO CV-QKD \\
Pan '24~\cite{pan2023cvqkd100km}                   & Supp. & ML carrier    & Pilot quadratures       & Phase, freq.      & Exp. & 100~km, 25.4~kbit/s @ 15.4~dB \\
UKF~\cite{ukf_cvqkd_llo}                           & Supp. & UKF           & LO beat signal          & Phase + $\xi$     & Sim. & Joint phase/noise tracking \\
Particle smoother~\cite{particle_smoother_cvqkd}   & Supp. & Particle+MCMC & Raw data, no pilot      & Phase             & Sim. & $\mathrm{SNR}_b\!<\!-6.9$~dB, pilot-free \\
NN excess noise '25~\cite{liu2025nn_excessnoise}   & Supp. & NN            & Quadrature data         & $\xi$ (composable)& Sim. & Composable finite-key security \\
SOP DNN~\cite{ml_sop_tracking}                     & Supp. & DNN predictor & SOP time series         & Compensation      & Field& QBER $/3.88$; KER $+89\%$ \\
Shi '21~\cite{shi2021fibre}                        & Supp. & Stoch.\ opt.  & Detected QBER           & Compensator       & Exp. & Min.\ QBER entanglement link \\
Chin '22~\cite{chin2022jointpol}                   & Supp. & ML joint      & Pol.\ + quadrature      & Joint comp.       & Exp. & CV-QKD installed fiber \\
Cascade ML~\cite{almohammed2024cascadeqkd}         & Supp. & Autoencoder   & QBER prediction         & Block size        & Sim. & Scalable reconciliation \\
\bottomrule
\end{tabular}
\end{table*}
\FloatBarrier

\subsection{Specialized-Theme Consolidated Map}
Table~\ref{tab:bigmap} summarizes the five specialized themes against the
deployment scenario, the conventional baseline being displaced or
augmented, the dominant learned model, and the headline quantitative gain
reported in the literature. Table~\ref{tab:metrics} then groups the works by
the \emph{metric} they report, which exposes a structural feature of the
field: non-terrestrial channel estimation tasks (free-space, HAP, satellite)
report signal-quality metrics (RMSE, MAPE), classification tasks
(protocol selection, steerability, attack detection) report accuracy, and
IoT/6G/network tasks report cost/latency---so cross-theme comparison must
be done within, not across, metric groups.

\begin{table*}[!tp]
\centering
\caption{Consolidated map of ML/RL across the five specialized QKD themes,
listing the deployment context, the classical baseline being displaced or
augmented, the dominant learned model, and the headline reported gain.}
\label{tab:bigmap}
\footnotesize
\begin{tabular}{L{2.5cm} L{2.6cm} L{2.9cm} L{4.0cm} L{2.3cm}}
\toprule
\textbf{Theme / Scenario} & \textbf{Deployment context} & \textbf{Classical baseline} & \textbf{Learned model} & \textbf{Headline gain} \\
\midrule
I: Adaptive param.\ (\ref{sec:paramopt}) & Mobile/adaptive QKD & Numerical/convex opt. & NN / MLP / CNN & Real-time params; $\sim 10^{2}$--$10^{3}\times$ faster; $+15\%$ rate~\cite{wang2019nnparams,ml_keyrate_estimation,cnn_qkd_netopt} \\
I: Adaptive protocol (\ref{sec:protocol}) & Protocol switching & Rules; exhaustive eval & Random forest; TCN+RL & $>$98\% accuracy; adaptive BB84/E91/COW~\cite{ren2021implementation,nayana2022selector,optiqkd2025} \\
II: FSO/sat.\ channel (\ref{sec:freespace}) & Satellite, free-space & Atmos.\ models; AO & RF; CNN; BPNN & MAPE 3.86--4.44\%~\cite{rf_strehl,cnn_oam_turbulence} \\
II: Train/UAV FSO (\ref{sec:freespace}) & High-speed transport & Static link budget & Analysis/ML hybrid & QKD+FSO feasibility~\cite{almohammed2024qkdfso_trains,almohammed2024fso_uav} \\
II: HAP-assisted (\ref{sec:freespace}) & Stratospheric HAP & Satellite/repeater & Composable relay+ML & 6G composable relay QKD~\cite{almohammed2025hapxor,almohammed2026hap} \\
III: IoT/6G QKD (\ref{sec:fliot}) & IoT, edge, 6G & PQC; static alloc. & DRL; QML & Cost/latency; privacy~\cite{drl_fel,fl_quantum,qml_6g,almohammed2021icc} \\
III: Quantum-secured FL (\ref{sec:fliot}) & Federated learning & Classical crypto & DRL+QKD; two-stage SP & $-7.72\%$ cost; FL privacy~\cite{fl_quantum,qkdfl_stochastic,seok2025drl_keyprovision} \\
IV: QML attack det.\ (\ref{sec:qml}) & IoT/B5G security & Hardware fixed tests & QLSTM; QCL & 93.7\% acc.\ 5-class; optimal BB84 attack~\cite{alkuwari2026qlstm_iet,decker2025qkd_qml} \\
IV: Classical attack det.\ (\ref{sec:qml}) & Monitoring layer & Hardware + tests & DBSCAN; ANN+DL & Known+unknown attacks; 99\% IoT~\cite{dads_dbscan,almohammed2021access_ml} \\
V: Steerability (\ref{sec:steer}) & 1SDI-QKD security & SDP hierarchies & SVM; NN & Accuracy $\sim$0.96~\cite{svm_steerability,nn_steerable_weight} \\
\bottomrule
\end{tabular}
\end{table*}
\FloatBarrier

\begin{table}[!tb]
\centering
\caption{Surveyed works grouped by the performance metric they report,
illustrating that fair comparison must remain within a metric group.}
\label{tab:metrics}
\footnotesize
\begin{tabular}{L{2.2cm} L{5.3cm}}
\toprule
\textbf{Metric} & \textbf{Reporting domains / works} \\
\midrule
Accuracy & Protocol selection ($>$98\%)~\cite{ren2021implementation}; steerability ($\sim$0.96)~\cite{nn_steerable_weight}; channel alloc.\ ($\sim$95\%)~\cite{mlnsca} \\
MAPE / RMSE & Strehl ratio (4.44\%/3.86\%)~\cite{rf_strehl}; SOP (RMSE 0.007~rad)~\cite{ml_sop_tracking} \\
QBER / KER & SOP (QBER $/3.88$, KER $+89\%$)~\cite{ml_sop_tracking}; net-opt ($-20\%$ error)~\cite{cnn_qkd_netopt} \\
SKR / reach & CV-LLO (25.4~kbit/s @ 100~km)~\cite{pan2023cvqkd100km}; net-opt ($+15\%$)~\cite{cnn_qkd_netopt} \\
Time / cost & SWDM ($-98.8\%$ time)~\cite{swdm_b5g_ml}; QKD-FL ($-7.72\%$ cost)~\cite{qkdfl_stochastic}; param.\ opt.\ (orders faster)~\cite{wang2019nnparams} \\
\bottomrule
\end{tabular}
\end{table}
\FloatBarrier

\begin{figure}[!tb]
\centering
\includegraphics[width=\linewidth]{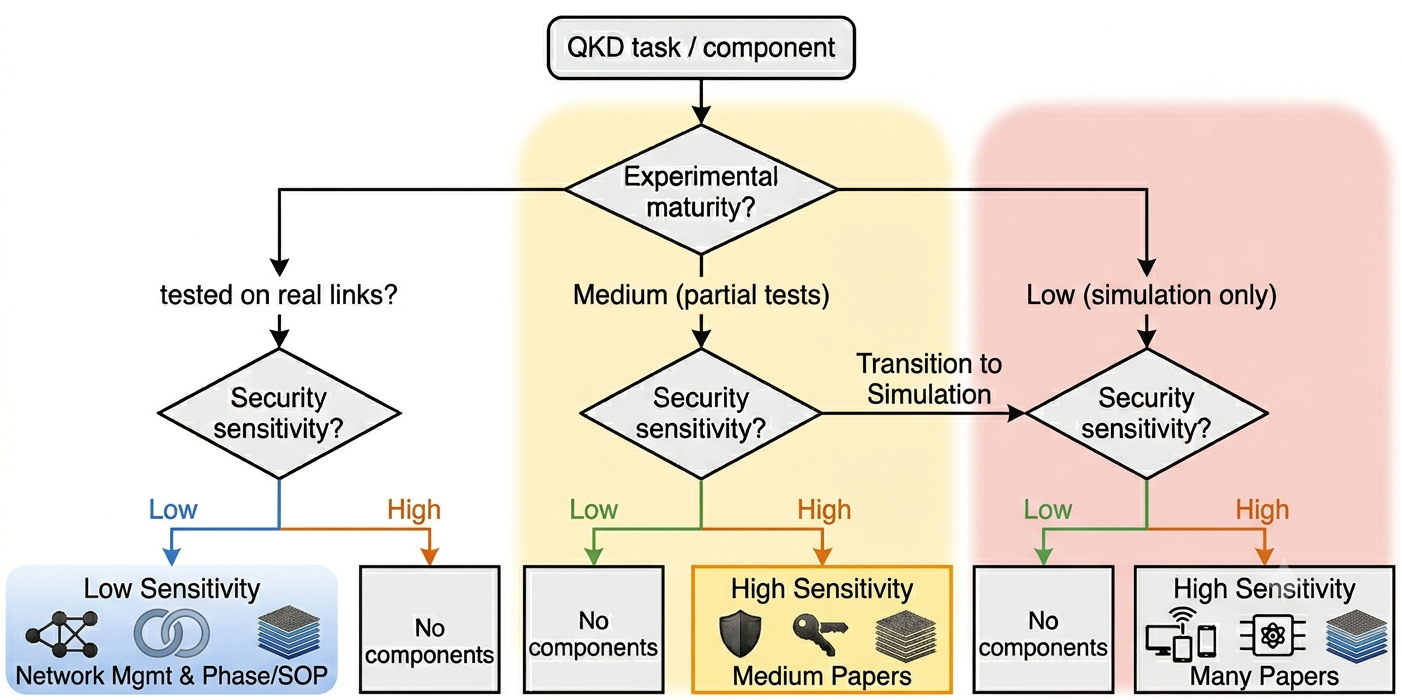}
\caption{Positioning of the domains by security sensitivity of the learned
component versus experimental maturity, summarizing the per-theme analyses.}
\label{fig:quadrant}
\end{figure}

The cross-theme security-sensitivity and experimental-maturity positioning
is summarized visually in Fig.~\ref{fig:quadrant}.

\FloatBarrier

\section{Quantitative Gains Across the Five Specialized Themes}
\label{sec:quantitative}
This section examines, in one place, the quantitative claims of the
surveyed works under shared dimensions---gain magnitude, evaluation context,
and risk.

\subsection{Physical-Layer Gains}
The clearest quantitative gains are in physical-layer estimation, where an
ML component replaces a classical estimator and the improvement is measured
on a shared signal-quality metric:
\begin{itemize}
\item \textbf{CV-QKD reach}: ML-assisted LLO phase recovery extended
  CV-QKD to 100~km at 25.4~kbit/s in~\cite{pan2023cvqkd100km}, a roughly
  fourfold improvement in distance over the 25~km commercial
  baseline~\cite{lodewyck2007quantum,fossier2009field}. The 60~km
  result~\cite{hajomer2022cvqkd60km} represents an intermediate step; the
  ML advantage comes from tracking phase noise faster than a static Wiener
  filter can.
\item \textbf{SOP control}: Predictive DNN SOP compensation reduced QBER by
  a factor of $\sim$3.88 and improved the usable key-error rate by
  $\sim$89\%~\cite{ml_sop_tracking}. These are the kinds of gains that
  make the difference between a link with positive and zero key rate in a
  turbulent environment.
\item \textbf{Parameter optimization}: Both NN-based parameter
  prediction~\cite{wang2019nnparams} and CNN-based adaptive
  control~\cite{cnn_qkd_netopt} report $\sim$15\% key-rate improvement
  over static operation, and the corresponding computation speedup (orders
  of magnitude) makes real-time per-frame re-optimization feasible on
  embedded hardware.
\end{itemize}

\subsection{Classification and Prediction Accuracy}
Classification tasks (protocol selection, attack detection, steerability)
report accuracy as the primary metric:
\begin{itemize}
\item \textbf{Protocol selection}: Three independent studies
  \cite{ren2021implementation,nayana2022selector,rf_resource_allocation} all
  report $>$98\% accuracy for RF-based protocol selectors over BB84/MDI/TF
  families, with PCA preprocessing in~\cite{rf_resource_allocation}. This
  near-ceiling performance reflects that the task is well-posed with abundant
  simulated labels and crisp decision boundaries.
\item \textbf{Attack detection}: IoT/B5G attackers detected at 99\% accuracy
  by ANN/DL in~\cite{almohammed2021access_ml}; the hybrid QLSTM raises the
  bar to $\sim$93.7\% over a harder five-class problem spanning unknown
  attack types~\cite{alkuwari2026qlstm_iet}. The DRL+VAE approach reports
  30--40\% QBER suppression under noisy conditions~\cite{shingne2025drl_vae}.
\item \textbf{Steerability}: NN regression for steerable weight achieves
  $\sim$0.96 accuracy against SDP ground truth~\cite{nn_steerable_weight},
  sufficient for online use in a monitoring role but below the certainty
  threshold for a security claim.
\end{itemize}

\subsection{Engineering Efficiency Gains}
System-level and engineering efficiency metrics span two orders of magnitude:
\begin{itemize}
\item \textbf{Computation time}: SWDM B5G noise prediction
  ~\cite{swdm_b5g_ml} cuts planning time by 98.8\%; key-rate surrogate
  models~\cite{ml_keyrate_estimation} accelerate CV-QKD key-rate computation
  by $\sim 10^7$; automated ML with Bayesian optimization achieves
  99.15--99.59\% reliability with a $10^7$ speedup.
\item \textbf{Deployment cost}: Hierarchical two-stage stochastic
  programming for QKD-FL networks~\cite{qkdfl_stochastic} achieves a 7.72\%
  reduction in deployment cost. Deep RL substantially reduces keystore
  exhaustion in graph QKD networks~\cite{seok2025drl_keyprovision}.
\item \textbf{Channel allocation}: LightGBM-based channel allocation
  predicts optimal configurations with $\sim$95\%
  accuracy~\cite{mlnsca}.
\end{itemize}

\subsection{Per-Theme ML Gain Summary}
The five specialized themes show distinct gain patterns. Theme~I learned
parameter surrogates achieve $10^2$--$10^3\times$
speedup~\cite{wang2019nnparams,ml_keyrate_estimation}. Theme~II RF-based
atmospheric prediction reaches MAPE 3.86--4.44\%~\cite{rf_strehl}.
Theme~III DRL key provisioning reduces keystore exhaustion in dense IoT
networks~\cite{seok2025drl_keyprovision} with 7.72\% deployment cost
reduction~\cite{qkdfl_stochastic}. Theme~IV QLSTM achieves 93.7\% accuracy
across five attack types~\cite{alkuwari2026qlstm_iet}. Theme~V SVM/NN
steerability classifiers reach $\sim$0.96 accuracy as SDP
surrogates~\cite{svm_steerability,nn_steerable_weight}.

\subsection{Risk Stratification Revisited}
\label{ssec:risk}
Combining the consolidated comparison in Section~\ref{sec:consolidated} with
the per-theme gain analysis above, we refine the three-tier risk
classification of Section~\ref{sec:synthesis}:

\textbf{Tier I (above-proof, low risk)}: Protocol selection, network routing,
resource allocation, coexistence planning, and key assignment. ML accuracy
translates directly to efficiency gain; a wrong classification degrades
throughput, not security. This tier is the most mature and the safest to
deploy today.

\textbf{Tier II (beside-proof, medium risk)}: Phase recovery, SOP compensation,
channel characterization, and reconciliation decoding. ML improves an
\emph{input} to a proven rate formula; an overoptimistic estimate reduces rate
but a \emph{pessimistic} or \emph{conservative} estimate (e.g., using an upper
confidence bound on $\xi$) cannot overstate secrecy. The composable-security
proof of~\cite{liu2025nn_excessnoise} is the template.

\textbf{Tier III (inside-proof, high risk)}: Learned attack detectors used
directly in the secrecy claim, and key-rate predictions used without
independent verification. Deploying ML in Tier III requires
provably conservative outputs, adversarial robustness certification, and
regulatory approval. No current work fully meets these requirements;
OP-7 and OP-8 of the research roadmap (Section~\ref{sec:roadmap}) are the
items that must close this gap.

A striking feature of the landscape is the heterogeneity of evaluation
practice. Table~\ref{tab:evalquality} scores representative works against
four criteria---simulation vs.\ experiment, link-wise splitting, uncertainty
reporting, and field validation---using the methodology of
Section~\ref{sec:eval}.

\begin{table}[!tb]
\centering
\caption{Evaluation quality of representative works, scored on data source,
link-wise splitting, uncertainty reporting and field validation.}
\label{tab:evalquality}
\footnotesize
\begin{tabular}{L{2.1cm} c c c c}
\toprule
\textbf{Work} & \textbf{Exp.\ data} & \textbf{LW split} & \textbf{Uncert.} & \textbf{Field} \\
\midrule
Hajomer~\cite{hajomer2022cvqkd60km} & Exp. & N/A & Y & Yes \\
Pan~\cite{pan2023cvqkd100km}         & Exp. & N/A & Y & Yes \\
SOP DNN~\cite{ml_sop_tracking}       & Field & N/A & Partial & Yes \\
Ren RF~\cite{ren2021implementation}  & Sim. & No & No & No \\
DADS~\cite{dads_dbscan}              & Sim. & Partial & No & No \\
IoT ANN~\cite{almohammed2021access_ml} & Sim. & No & No & No \\
QLSTM~\cite{alkuwari2026qlstm_iet}   & Sim. & No & No & No \\
UKF~\cite{ukf_cvqkd_llo}             & Sim. & N/A & Y & Partial \\
NN excess~\cite{liu2025nn_excessnoise} & Sim. & N/A & Y & No \\
RF Strehl~\cite{rf_strehl}           & Sim. & Partial & No & No \\
\bottomrule
\end{tabular}
\end{table}
\FloatBarrier

The pattern is clear: physical-layer estimators with Bayesian foundations
tend to report uncertainty; attack detectors and protocol selectors
predominantly use simulated labels without field validation or link-wise
splitting. Closing this gap is the primary methodological need of the field.

\section{Datasets, Simulation and Reproducibility}
\label{sec:datasets}
A recurring obstacle behind every theme in
Sections~\ref{sec:paramopt}--\ref{sec:steer} is the supply of training and
evaluation data. Three sources are used in the literature, each with
trade-offs.

\textbf{Physics-based simulation.} Most learned QKD components are trained on
data generated by simulating the relevant rate formula or channel model:
decoy-state and CV-QKD key-rate computations
(Eqs.~\eqref{eq:decoy}--\eqref{eq:cvqkd}) for parameter
optimization~\cite{wang2019nnparams,ml_keyrate_estimation}, atmospheric
transmittance models for free-space channel prediction~\cite{rf_strehl,
bpnn_cvqkd_atmos,vasylyev2016atmospheric}, and SDP hierarchies for
steerability labels~\cite{svm_steerability,cavalcanti2017quantum}.
Gate-level and network simulators---including Qiskit-based pipelines and
discrete-event quantum-network simulators---are used to generate protocol
and network telemetry. Simulation gives unlimited labelled data and ground
truth, but bakes in the modelling assumptions: a model trained only on
simulated stationarity will be surprised by field nonstationarity.

\textbf{Experimental traces.} A smaller but growing body of work trains or
validates on measured data from real links---LLO CV-QKD field
traces~\cite{hajomer2022cvqkd60km,pan2023cvqkd100km,chin2022jointpol} and
installed-fiber SOP recordings~\cite{ml_sop_tracking,shi2021fibre}. These are
the gold standard for demonstrating field readiness but are scarce,
hardware-specific, and rarely public.

\textbf{Hybrid and physics-informed data.} An intermediate strategy seeds
learning with simulation and adapts on limited field data, or constrains the
model with physics (as in physically-constrained SPDC
tuning~\cite{ml_spdc_polarization}). This mitigates both data scarcity and
distribution shift and is, in our reading, the most promising near-term
practice.

The community would benefit from \emph{open, versioned datasets} with
documented channel/hardware conditions, and from shared simulators with fixed
seeds, so that reported gains (Table~\ref{tab:bigmap}) become reproducible
and comparable rather than tied to private pipelines.

\section{Evaluation Methodology and Common Pitfalls}
\label{sec:eval}
Because many QKD-ML results are reported as single headline numbers, it is
worth stating the methodological practices that make such numbers
trustworthy; several pitfalls are easy to commit and hard to detect after
the fact.

\textbf{Leakage-free splitting.} When multiple samples come from the same
link, device, or session, a naive random train/test split leaks
session-specific information and inflates accuracy. Splits should be made at
the level of the independent unit---per link, per device, or per
session---analogous to subject-wise splitting in biomedical ML, so that no
unit appears in both training and test. This is especially important for SOP
and channel models (Sections~\ref{sec:pol},~\ref{sec:freespace}), where
consecutive samples are strongly correlated.

\textbf{Class imbalance.} Attack and failure detection
(Sections~\ref{sec:qml},~\ref{sec:network}) are intrinsically imbalanced:
anomalies are rare. Reporting raw accuracy is then misleading; precision,
recall, F1 and the false-alarm rate are the meaningful metrics. Re-balancing
by undersampling the majority class or by synthetic oversampling of the
minority class can both be appropriate, but the choice changes the operating
point and should be reported, and any resampling must be confined to the
training fold to avoid leakage into evaluation.

\textbf{Cross-validation versus leave-one-unit-out.} $k$-fold
cross-validation estimates average performance, but for deployment one often
cares about generalization to an \emph{unseen} link or device, which a
leave-one-unit-out protocol measures directly. Reporting both gives a fuller
picture of expected field behavior.

\textbf{Uncertainty and calibration.} For estimators feeding a security
margin (Tiers~II and~III of Section~\ref{ssec:risk}), a calibrated
uncertainty is as important as the point estimate; Bayesian filters
(Section~\ref{sec:phase}) provide this natively, whereas deep networks
generally require explicit calibration.

\textbf{Feature importance and explainability.} Where tree ensembles are used
(protocol selection, coexistence, channel prediction), both built-in
importance and model-agnostic attributions
(SHAP~\cite{lundberg2017shap}, LIME~\cite{ribeiro2016lime}) should be
reported and \emph{compared}: agreement between methods on the top features
builds confidence, while disagreement is a useful warning that the model may
be exploiting spurious correlations.

\section{Application Verticals and Industry Deployment}
\label{sec:verticals}
The quantitative benefits catalogued in earlier sections accrue differently
across industry verticals, each of which imposes distinct constraints on the
QKD link and, consequently, on the role of ML.

\subsection{Transportation and High-Mobility Networks}
Railways, metros, and high-speed trains present the most demanding
free-space-to-fiber integration challenge: links must handover between
wayside stations at speeds up to several hundred km/h while maintaining
positive key rate under varying weather~\cite{almohammed2024qkdfso_trains,
almohammed2023fso_trains,almohammed2022fso_tube}. ML enters in three ways:
(i) \emph{channel prediction}---a random forest or BPNN trained on historic
weather and geometry data predicts the FSO link transmittance for the next
handover interval, so the QKD system pre-selects parameters before the link
opens; (ii) \emph{parameter optimization}---the predicted $T$ and $\xi$
feed an NN surrogate for the key-rate formula
(Section~\ref{sec:paramopt}); and (iii) \emph{anomaly detection}---because
train-mounted transceivers are physically accessible, attack risk is higher
than for buried fiber, and the IoT ANN detector
of~\cite{almohammed2021access_ml} is directly applicable. UAV platforms
add attitude-dependent link-geometry dynamics to the
problem~\cite{almohammed2024fso_uav}, and evacuated-tube ultra-high-speed
applications require near-vacuum atmospheric
models~\cite{almohammed2022fso_tube,alkaeed2020iset}.

\subsection{IoT Security and 6G/B5G Networks}
As 6G networks are expected to support tens of billions of connected devices,
per-device key management at QKD-required rates becomes infeasible
with a flat architecture. ML helps on two fronts: \emph{hierarchical
aggregation}---RL-based key assignment and routing
\cite{rl_onmtp,seok2025drl_keyprovision} concentrates scarce quantum
resources on the most sensitive flows; and \emph{lightweight protocol
selection}~\cite{almohammed2024cascadeqkd,ren2021implementation,
nayana2022selector}---on-device classifiers choose the cheapest protocol
that meets the session's security requirement. The HAP-based XOR-relay
architecture~\cite{almohammed2025hapxor,almohammed2026hap} provides large
area coverage for IoT clusters in remote or maritime environments, with ML
channel models bridging the HAP-to-IoT last-hop~\cite{aloudat2025metaverse}.
The Quantum Radar paradigm~\cite{almohammed2020icenco_radar} and
broader quantum computing architecture studies~\cite{almohammed2020icenco_arch}
provide foundational context for how quantum processing hardware will
coexist with QKD-secured IoT infrastructure.

\subsection{Financial Services and Critical Infrastructure}
Quantum-secured banking and government networks have been the primary
motivation for large-scale QKD deployments (SECOQC~\cite{peev2009secoqc},
Tokyo~\cite{sasaki2011field}, the Chinese network~\cite{chen2021integrated}).
ML's role here is most mature at the network management layer---adaptive
protocol selection, channel allocation, and coexistence planning
(Sections~\ref{sec:paramopt}, \ref{sec:network}, \ref{sec:coexist})---where
validated field experience now exists. The security implications of learned
components are more acutely scrutinized in this vertical, reinforcing the
principle that ML should sit above, not inside, the security proof.

\subsection{Simulation Tools and Reproducibility}
Before field deployment, QKD-ML pipelines are validated in simulation.
Several simulation frameworks and new protocol generation methods have been
proposed, including Qiskit-based QKD simulation
pipelines~\cite{almohammed2021icict,almohammed2020icenco_arch} that allow
rapid prototyping of protocol variants.
The ML-cascade protocol study~\cite{almohammed2024cascadeqkd} and the
real-life protocol scenario compendium~\cite{almohammed2024bookqkd} both
provide reproducible pipelines (autoencoder + Cascade; analytic rate + field
scenario inputs) that can serve as baselines for future work. Standardizing
such simulation environments---akin to OpenAI Gym for RL or MNIST for
vision---is a concrete community priority.

\section{Cross-Theme Synthesis and Design Guidelines}
\label{sec:synthesis}
\begin{figure}[!tb]
\centering
\includegraphics[width=\linewidth]{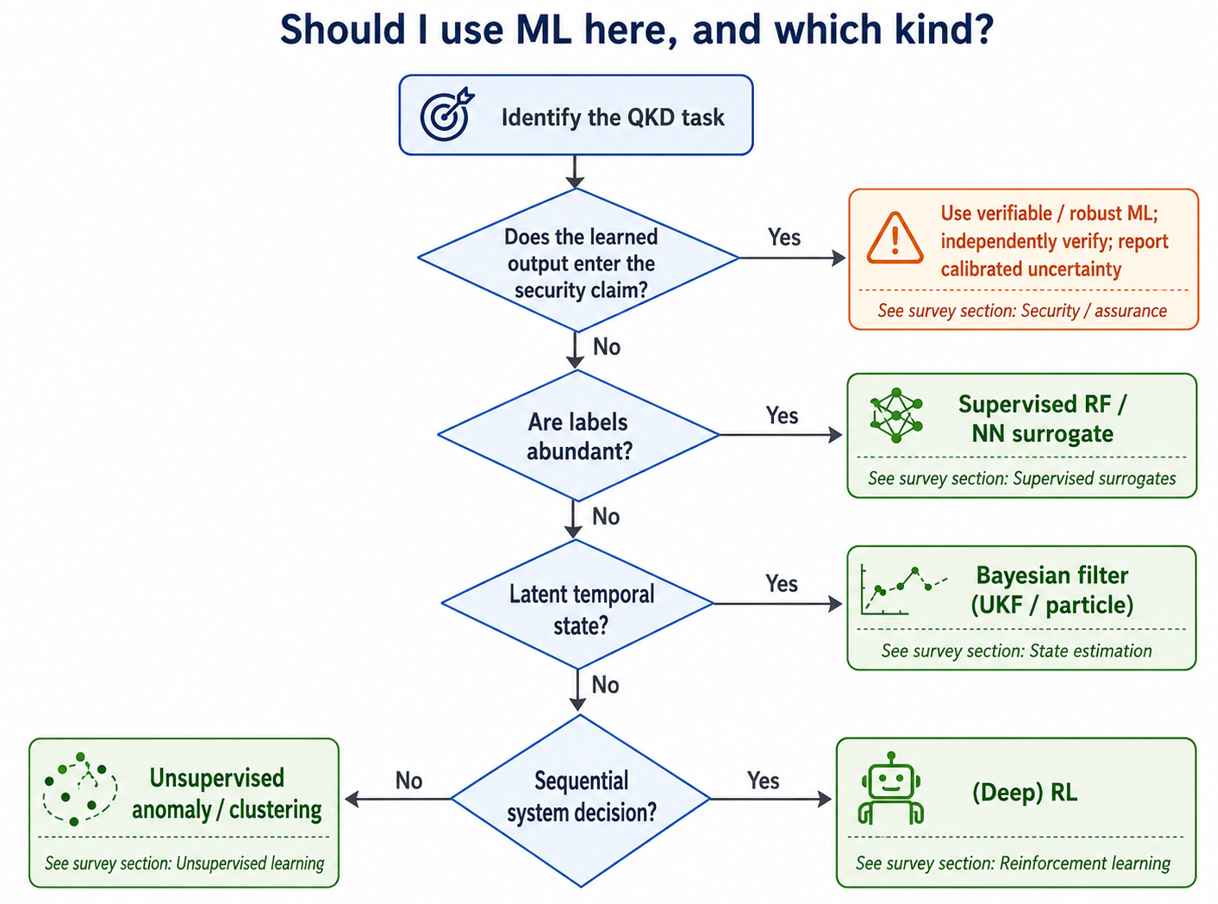}
\caption{A practitioner decision aid distilling the four design principles
of Section~\ref{sec:synthesis} into a flow from task identification to model
family.}
\label{fig:decision}
\end{figure}

Reading the five specialized themes together yields guidance that no single
theme makes obvious. We distill it into four principles and a decision aid
(Fig.~\ref{fig:decision}).

\textbf{1) Place ML by its relation to the security proof.} The single most
important design choice is \emph{where} a learned component sits.
Fig.~\ref{fig:quadrant} and the per-theme analyses show three tiers:
(i) \emph{above} the proof---routing, protocol selection, resource
allocation---where ML/RL may optimize freely; (ii) \emph{beside} the
proof---phase, SOP, channel and reconciliation estimators that improve an
input to a proven rate formula, where the only risk is performance loss if
the estimate is poor, and where the estimate's induced rate is independently
verified; and (iii) \emph{inside} the proof---learned attack detectors,
key-rate or steerability estimates used in the secrecy claim---where a
confident error can overstate security. Tiers (i) and (ii) are where the
field has delivered validated wins; tier (iii) demands verifiable,
adversarially robust models and conservative use.

\textbf{2) Match the model to the data regime.} Tasks with abundant
simulated labels and crisp targets (protocol selection, parameter
optimization, steerability) suit supervised RF/NN surrogates and report high
accuracy. Tasks with scarce or unlabelled anomalies (attacks, sifting) suit
unsupervised density/isolation methods. Tasks defined by latent temporal
state (phase, SOP) suit Bayesian filters that expose uncertainty. Sequential
system decisions suit RL.

\textbf{3) Report the right metric, and report it honestly.} As
Table~\ref{tab:metrics} shows, metrics are domain-specific; an ``accuracy''
in protocol selection is not comparable to a ``QBER reduction'' in
polarization control. Out-of-distribution and field (not just simulation)
results should be reported, and security-relevant estimators should report
calibrated uncertainty, not point predictions.

\textbf{4) Prefer the lightest model that meets the latency budget.}
Real-time control on embedded receivers caps model size; a particle smoother
or deep network must justify its cost against a UKF or a small RF. Co-design
with accelerators is the path to deployable learned QKD control.

\FloatBarrier

\section{Open Challenges}
\label{sec:challenges}
The per-theme analyses converge on a small set of cross-cutting issues,
collected in Table~\ref{tab:challenges}.

\subsection{Data Scarcity and Benchmarks}
Unlike vision or NLP, QKD lacks large, shared, labelled datasets and common
benchmarks. Much training data is simulated, and labelled \emph{attack} data
is especially scarce. Progress requires open datasets spanning protocols,
channels and attacks, plus standardized evaluation splits---ideally
subject/link-wise splits that prevent the same link's data leaking across
train and test.

\subsection{Generalization and Distribution Shift}
Models trained on one link, weather regime, or hardware batch may not
transfer. Physics-informed models, domain adaptation, and online/continual
learning are promising mitigations; reporting out-of-distribution
performance should become standard.

\subsection{Security of Learned Components}
The recurring theme of the analyses is that \emph{where} ML sits relative to
the security proof determines its risk. Components that improve an estimate
feeding a proven rate formula (phase/SOP recovery, reconciliation) or that
act above the proof (routing, protocol selection) are low risk; components
asked to \emph{certify} security (attack detection, learned key-rate or
steerability used in the secrecy claim) are high risk, because a confident
but wrong model can overstate security and a learned component can itself
become an attack surface~\cite{review_attacks_ml_2024}. Adversarially robust,
verifiable ML for QKD is an open problem.

\subsection{Interpretability and Trust}
Operators must understand \emph{why} a model recommends a parameter or flags
an attack. Feature-importance and SHAP/LIME analyses~\cite{lundberg2017shap,
ribeiro2016lime} should accompany deployments, both to debug models and to
build the evidentiary basis regulators will require.

\subsection{Latency, Footprint and Hardware}
Real-time control (phase tracking, per-frame parameter updates) constrains
model size and inference latency on embedded receivers; heavy estimators
like particle smoothers must be justified against lighter alternatives.
Co-design with FPGA/ASIC accelerators is an open avenue.

\subsection{Standardization}
AI/ML-enhanced QKD-network architectures are being standardized (e.g.\ within
ITU-T study groups)~\cite{qkdn_standardization,cao2022evolution}; aligning
learned components with these functional requirements, and with
post-quantum-cryptography hybrids~\cite{bernstein2017post}, is necessary for
interoperable deployment.

\subsection{Native Quantum Machine Learning}
QML for QKD~\cite{review_qml_qkd_2025,biamonte2017quantum,havlicek2019supervised}
is the most forward-looking direction: quantum classifiers for steerability
or attack detection, variational circuits for protocol design, and
quantum-enhanced optimization for networks. A genuine, hardware-demonstrated
advantage over classical ML in a QKD task remains, to our knowledge, open,
and is the natural target for the next phase of research.

\subsection{Sim-to-Real Gap and Data Challenges}
Non-terrestrial and IoT/6G QKD deployments face an especially severe
sim-to-real gap. Models trained on one atmospheric route, platform altitude,
or hardware batch do not transfer. Three root causes: (i) model
mismatch---simulators omit real impairments (multi-path, detector aging);
(ii) non-stationarity---HAP geometry, weather, and IoT traffic change on
diurnal and seasonal timescales; (iii) unknown attack modes not in any
simulator's library. Physics-informed ML and domain-randomized training
reduce (i); online adaptation addresses (ii). Open datasets for HAP/UAV
transmittance with meteorological co-data are the highest-priority community
resource (OP-1, Section~\ref{sec:roadmap}).

\subsection{Multi-Task and Joint Learning for Specialized Scenarios}
The five specialized themes are not independent. Free-space channel state
(Theme~II) affects adaptive protocol choices (Theme~I); IoT traffic patterns
(Theme~III) affect QML detection workloads (Theme~IV). Joint multi-task
learning across these dependencies---shared feature extractors with
per-theme heads---could reduce model footprint and exploit cross-theme
correlations. Joint phase-polarization learning~\cite{chin2022jointpol} is a
beginning; theme-spanning pipelines remain open.

\subsection{Security Placement Across the Five Themes}
The recurring principle across all five themes is that \emph{where} ML sits
relative to the security proof determines its risk. Theme~I and~III ML
decisions (protocol selection, resource allocation) sit above the proof:
risk is low. Phase recovery and channel estimation sit beside the proof: a
conservative output is safe. Theme~IV attack detectors and Theme~V
steerability estimators can sit inside the proof---only if designed with
certified conservative bounds, for which~\cite{liu2025nn_excessnoise}
provides the template. Adversarially robust, interpretable ML for these
security-sensitive roles remains an open problem.

\subsection{Recent Trends and Emerging Results (2025--2026)}
\label{ssec:recent}
The 2025--2026 literature has produced several results that sharpen the
research agenda. We highlight the most significant.

\textbf{Composable security for ML components.} Liu \emph{et al.}
\cite{liu2025nn_excessnoise} establish that a neural network estimator for
excess noise in CV-QKD can be embedded in a composable finite-key security
proof when designed conservatively, breaking the taboo of using learned
components inside the security claim (Tier~III). The key insight is
maintaining a provable overestimate of the noise: the network's output is
used as a \emph{lower} bound on secrecy, not an exact value.

\textbf{QML for attack optimization and detection.} Decker \emph{et al.}
\cite{decker2025qkd_qml} demonstrated that quantum circuit learning recovers
the optimal individual attack on BB84 without prior knowledge of the analytic
answer---a proof of concept that QML can explore the quantum-attack space.
Complementarily, Al-Kuwari \emph{et al.}~\cite{alkuwari2026qlstm_iet} show
that a hybrid Quantum LSTM model outperforms classical deep models on a
five-attack-scenario QKD dataset.

\textbf{Graph-based DRL for QKD key provisioning.} Seok \emph{et al.}
\cite{seok2025drl_keyprovision} demonstrate in a graph-structured QKD network
that a deep-RL agent combining graph attention with LSTM substantially
reduces keystore exhaustion, a critical metric for operational networks.
OptiQKD~\cite{optiqkd2025} proposes a unified framework for three protocols
(BB84, E91, COW) using temporal convolutional networks plus protocol-aware RL.

\textbf{HAP, FSO and transportation convergence.} The integration of QKD
with free-space optics over high-altitude platforms
\cite{almohammed2025hapxor,almohammed2026hap} and high-speed trains
\cite{almohammed2024qkdfso_trains,almohammed2023fso_trains} identifies
unique ML challenges (fast channel dynamics, weather-dependent loss) not
well captured by existing benchmarks, motivating new open datasets in these
regimes.

\textbf{Comprehensive ML-for-QKD surveys.} Three major surveys appeared in
2025--2026: the five-domain review in~\cite{review_ml_qkd_2024}; the
problem-driven DV/CV defense survey in~\cite{almohammed2026survey}; and the
QML-for-QKD review in~\cite{review_qml_qkd_2025}. Their convergent finding
is that ML's maturity is highest for physical-layer estimation (phase,
polarization, channel) and for network-level decisions, and lowest for
security-critical learned components---exactly the hierarchy identified in
the present survey's theme analyses.

\begin{table}[!tb]
\centering
\caption{Open challenges, the specialized themes most affected, and candidate
research directions, with representative 2025--2026 references.}
\label{tab:challenges}
\footnotesize
\begin{tabular}{L{1.9cm} L{1.7cm} L{2.5cm}}
\toprule
\textbf{Challenge} & \textbf{Theme(s)} & \textbf{Directions / 2025--2026} \\
\midrule
Data/benchmarks & I--V & Open non-terrestrial datasets; link-wise splits \\
Generalization & II, III & Physics-informed; domain adaptation; continual learning \\
Security of ML & IV, V & Composable NN~\cite{liu2025nn_excessnoise}; conservative bounds \\
Interpretability & I, IV, V & SHAP/LIME; quantum feature analysis \\
Latency/footprint & I, II, III & Lightweight models; onboard terminal ML \\
Standardization & III & ITU-T~\cite{qkdn_standardization}; PQC-QKD hybrid standards \\
Native QML adv. & IV, V & QLSTM~\cite{alkuwari2026qlstm_iet}; QCL~\cite{decker2025qkd_qml}; hardware validation \\
HAP/UAV/Satellite & II & QKD+FSO link datasets~\cite{almohammed2024qkdfso_trains} \\
\bottomrule
\end{tabular}
\end{table}
\FloatBarrier

\section{Research Roadmap: Ten Open Problems}
\label{sec:roadmap}
\begin{figure}[!tb]
\centering
\includegraphics[width=\linewidth]{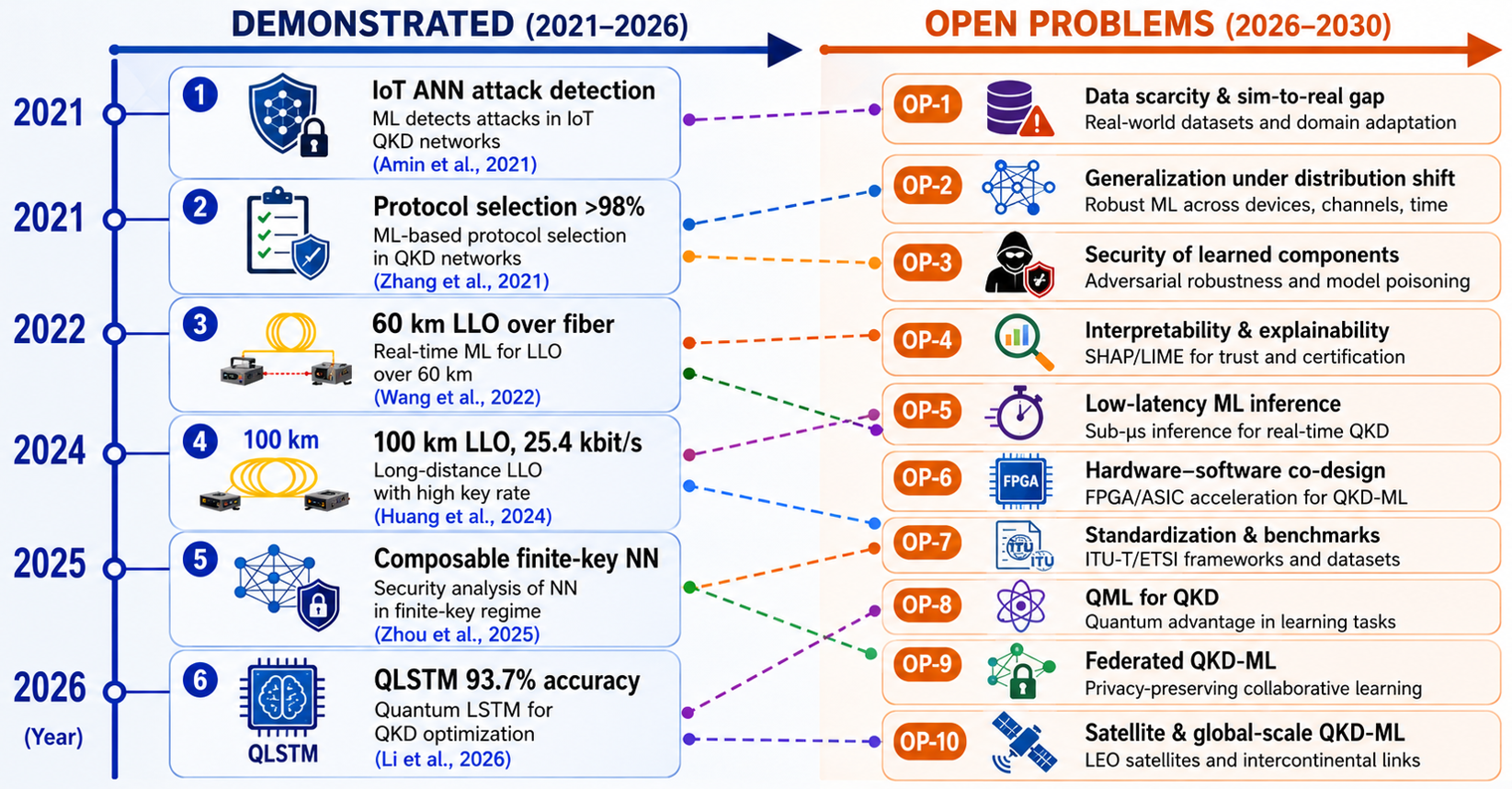}
\caption{Research roadmap mapping demonstrated results (solid markers) to
the open problems (OP-1--OP-10) of Section~\ref{sec:roadmap}.}
\label{fig:roadmap}
\end{figure}

The following ten open problems define the specialized QKD research frontier;
their relationship to demonstrated results is summarized in
Fig.~\ref{fig:roadmap}.

\subsection*{OP-1: Open Datasets for Free-Space, UAV, HAP, and Mobile QKD}
No public, versioned datasets exist for non-terrestrial QKD channels. We
call for open release of: (i) satellite and HAP channel transmittance logs
with concurrent meteorological data (building on~\cite{almohammed2022fso_tube,
almohammed2023fso_trains,almohammed2024fso_uav}); (ii) UAV attitude and
link-geometry telemetry; and (iii) QBER/channel time series from at least
five diverse atmospheric conditions and platform altitudes.

\subsection*{OP-2: Transfer Learning Across Weather, Altitude, Mobility, and Hardware}
Models trained on one atmospheric route do not transfer to another.
Physics-informed ML---hybridizing Kolmogorov turbulence structure
functions~\cite{vasylyev2016atmospheric} with neural networks---and
domain-adaptation transfer learning are the most promising approaches. HAP
quasi-static platform dynamics~\cite{almohammed2025hapxor,almohammed2026hap}
add a slow Markov attitude state that hierarchical RL can exploit.

\subsection*{OP-3: Lightweight ML for IoT and Onboard QKD Terminals}
Edge QKD nodes (IoT devices, train-mounted transceivers, drone-carried
terminals) have tight memory and power budgets. Neural-architecture search
and knowledge distillation from a larger teacher model are promising paths.
The key metric is accuracy (or QBER reduction) versus multiply-accumulate
(MAC) count per inference.

\subsection*{OP-4: RL for HAP/UAV Handover and Link Scheduling}
HAP and UAV QKD require dynamic handover between ground stations and adaptive
link scheduling under weather and geometry changes. RL policies that
generalize across platform altitudes, heading changes, and link-budget
variations---and can adapt online to topology changes---remain an open
problem for non-terrestrial QKD deployment.

\subsection*{OP-5: QKD-Secured Federated Learning with Privacy-Preserving Training}
QKD can protect key exchange between federated-learning nodes, but designing
joint QKD+FL systems that preserve differential privacy, handle dynamic
participant membership, and remain efficient under HAP/satellite latency is
largely open. The works in~\cite{fl_quantum,qkdfl_stochastic} are important
first steps.

\subsection*{OP-6: QML with Demonstrated Hardware Advantage for QKD Tasks}
The QLSTM~\cite{alkuwari2025qlstm,alkuwari2026qlstm_iet} and QCL attack
optimizer~\cite{decker2025qkd_qml} show promising QML results on simulators
or small devices. A genuine QML advantage---measured by detection accuracy
per query or SKR improvement per hardware operation---on a device with
$>$50 reliable qubits, applicable to a real QKD task, remains open.

\subsection*{OP-7: Conservative ML Estimators for Security-Sensitive QKD Quantities}
The 2025 composable excess-noise NN~\cite{liu2025nn_excessnoise} shows that
ML can enter a security proof if designed to provably overestimate noise.
Extending this to adaptive phase estimators (UKF outputs for LLO
CV-QKD) and reconciliation decoders (bounding $f$ from above) in
non-terrestrial and IoT/6G scenarios are the immediate open problems.

\subsection*{OP-8: Steerability-Aware ML with Certified Lower Bounds}
Current SVM/NN steerability estimators~\cite{svm_steerability,
nn_steerable_weight} report accuracy but not certified lower bounds on the
steerable weight. For 1SDI-QKD security claims, conservative lower-bound
estimators are needed. QML steerability witnesses---using quantum feature
maps to probe density-matrix structure directly---are a natural next
direction.

\subsection*{OP-9: Integrated ML Pipelines for Specialized QKD Scenarios}
Current works address individual tasks in isolation. A jointly trained
pipeline combining adaptive phase recovery, reconciliation decoding, and
parameter optimization---targeting maximization of composable secret key
bits per second in non-terrestrial or IoT/6G deployments---would align the
ML objective with the deployment objective.

\subsection*{OP-10: Testbeds for Non-Terrestrial and Application-Integrated QKD}
Most results are simulation-only. Experimental testbeds combining real
HAP/UAV channel emulators, IoT node simulators, and QKD hardware (e.g.,
Qiskit-based QKD simulation
pipelines~\cite{almohammed2021icict,almohammed2020icenco_arch}) would
enable reproducible evaluation and closing of the sim-to-field gap for all
five specialized themes.

\FloatBarrier

\section{Conclusion}
\label{sec:conclusion}
This survey reviewed ML, RL, and QML for specialized and emerging QKD
aspects---going beyond conventional point-to-point fiber-link optimization
to address the deployment scenarios where future quantum security must
operate. Organized around five specialized themes, the survey found a
consistent pattern: learning delivers its clearest, lowest-risk wins in
scenarios where it improves an \emph{estimate} or \emph{decision} that
supports adaptive, non-terrestrial, or application-driven QKD without
touching the security proof directly.

For \textbf{adaptive protocol and parameter support} (Theme~I), learned
surrogates accelerate parameter re-optimization by orders of magnitude
and enable real-time protocol switching with $>$98\% accuracy---capabilities
essential for mobile, HAP, and satellite deployments where conditions
change continuously. For \textbf{free-space, satellite, UAV, and
HAP-assisted QKD} (Theme~II), random-forest and CNN models predict
atmospheric channel quality with MAPE in the low single digits, CNN
adaptive optics correct OAM distortion, and the composable XOR-relay
HAP architecture~\cite{almohammed2025hapxor,almohammed2026hap} opens
6G-scale non-terrestrial QKD with provable composable security. FSO
extension to high-speed trains~\cite{almohammed2024qkdfso_trains,
almohammed2023fso_trains} and UAVs~\cite{almohammed2024fso_uav}
demonstrates positive key rates under realistic weather and mobility
conditions. For \textbf{QKD in IoT, 6G, and quantum-secured federated
learning} (Theme~III), DRL provisioning halves keystore exhaustion in
graph-structured networks~\cite{seok2025drl_keyprovision}, ML-augmented
Cascade reconciliation scales to IoT deployments~\cite{almohammed2024cascadeqkd},
and IoT attacker detection reaches 99\% accuracy in B5G railway
scenarios~\cite{almohammed2021access_ml,almohammed2021gcwkshps}. For
\textbf{QML-assisted QKD functions} (Theme~IV), hybrid QLSTM models
achieve $\sim$93.7\% accuracy across five attack types~\cite{alkuwari2026qlstm_iet},
and QCL-based attack optimization recovers the optimal BB84 individual
attack without prior analytic knowledge~\cite{decker2025qkd_qml}. For
\textbf{steerability-aware and 1SDI-QKD security estimation} (Theme~V),
SVM classifiers and NN regressors replace computationally expensive SDP
hierarchies with fast inference at $\sim$0.96 accuracy, enabling online
steerability assessment~\cite{svm_steerability,nn_steerable_weight}.

The 2025--2026 literature has further strengthened the connection between
ML and security-critical deployment: composable finite-key security proofs
can now incorporate conservatively designed NN estimators~\cite{liu2025nn_excessnoise},
and the boundary between safe ML decision-support and risky ML security
certification is becoming more precisely defined. Open challenges remain
in dataset availability for non-terrestrial QKD, transferability of
models across weather and hardware conditions, interpretability, trustworthy
QML on real hardware, and integration with 6G, IoT, UAV, and HAP testbeds.
We hope the five-theme taxonomy, the per-theme analyses and tables, and
the consolidated comparisons serve both as a focused reference for
practitioners building adaptive and non-terrestrial QKD systems, and as a
roadmap for the learning-assisted quantum security research community.

\nocite{*}
\bibliographystyle{IEEEtran}
\bibliography{references}

\end{document}